\documentclass[aps,10pt,showpacs,floatfix,superscriptaddress,twocolumn]{revtex4-2}
\usepackage{amsmath,amssymb,amsthm}
\usepackage{newtxtext,newtxmath}   
\usepackage[utf8]{inputenc}
\usepackage[T1]{fontenc}
\usepackage[english]{babel}
\usepackage{mathtools}
\usepackage{graphicx}
\usepackage{xcolor}
\usepackage{csquotes}\MakeOuterQuote"
\usepackage{float}
\usepackage{bbold}
\usepackage[caption=false]{subfig}
\usepackage{cases}
\usepackage[colorlinks]{hyperref}

\newcommand{\blue}[1]{{\color{blue}#1}}
\newtheorem{theorem}{Theorem}
\newcommand{\hide}[1]{}

\begin{document}

\title{ 
{Exact and lower bounds for the quantum speed limit in finite dimensional systems} 
}

\author{Mattias T. Johnsson }	
\affiliation{School of Mathematical and Physical Sciences, Macquarie University, North Ryde, NSW 2109, Australia}

\author {Lauritz van Luijk}
\affiliation{Institut f\"ur Theoretische Physik, Leibniz Universit\"at Hannover, Appelstrasse 2, 30167 Hannover, Germany}

\author{Daniel Burgarth}	
\affiliation{School of Mathematical and Physical Sciences, Macquarie University, North Ryde, NSW 2109, Australia}
\affiliation{Physics Department, Friedrich-Alexander Universität of Erlangen-Nuremberg, Staudtstr. 7, 91058 Erlangen, Germany}

\begin{abstract}
A fundamental problem in quantum engineering is determining the lowest time required to ensure that all possible unitaries can be generated with the tools available, which is one of a number of possible {\emph{quantum speed limits}}. We examine this problem from the perspective of quantum control, where the system of interest is described by a drift Hamiltonian and set of control Hamiltonians. Our approach uses a combination of Lie algebra theory, Lie groups and differential geometry, and formulates the problem in terms of geodesics on a differentiable manifold. We provide explicit lower bounds on the quantum speed limit for the case of an arbitrary drift, requiring only that the control Hamiltonians generate a topologically closed subgroup of the full unitary group, and formulate criteria as to when our expression for the speed limit is exact and not merely a lower bound. These analytic results are then tested and confirmed using a numerical optimization scheme. Finally we extend the analysis to find a lower bound on the quantum speed limit in the common case where the system is described by a drift Hamiltonian and a single control Hamiltonian.
\end{abstract}

\maketitle

\section{Introduction}

The emergence of quantum technologies such as quantum information processing, quantum engineering and quantum sensing has relied on our increasing ability to manipulate quantum systems with high levels of precision. Such manipulation requires the ability to carry out quantum operations and state preparation with high fidelity, in the presence of noisy environments, as quickly as possible, and potentially subject to a number of real-world constraints. 

These requirements are the province of the field of quantum control, which is primarily concerned with methods of steering a quantum system using a set of classical control inputs to the system \cite{glaser2015, brif2010}. Two major topics within this field are characterising the operations that can be carried out and the states that can be reached with a given set of controls, as well as determining the specific time dependence of those controls that will steer the system to the intended goal. The questions regarding the gates that can be implemented and state reachability are approached using the methods of bilinear control theory \cite{dalessandro2008,elliot2009} which usually involve a Lie theoretic framework \cite{dalessandro2008,jurdjevic1972,jurdjevic1996}. The questions regarding the determination of the time-dependent control fields (pulses) on the other hand, have no good general strategy and are generally difficult. Analytic methods of optimal control theory can be employed \cite{dalessandro2008,sontag1998,bechhoefer2021}, but usually numerical optimization is used, typically involving gradient-based search strategies with some fidelity cost functional \cite{khaneja2005,machnes2011,floethe2012,sakai2019}.

While these aspects of quantum control have been extensively studied, less attention has been given to the question of the speed at which specific unitaries can be generated or specific states can be reached. Given that decoherence is present in all quantum information processing, it is important to minimize the time taken to perform quantum operations. The time taken to reach specific targets given the set of controls available is known as the quantum speed limit \cite{cavena2009, deffner2017, lee2018}. Or, more precisely, there are a number of different speed limits, some for the transformation of states, some for unitary transformation, some for uncontrolled dynamics, and some for controlled dynamics \cite{deffner2017}.

We will be more precise later, but in general terms the quantum speed limit we will consider in this paper is the following: Assuming we have a set of controls that allow us to achieve all possible unitaries in a finite-dimension system, what is the minimum time by which we can guarantee we can produce all possible unitaries? That is, how much time must we allow to be certain that we can accomplish everything that can be done with the system?

The exact time for this type of quantum speed limit is generally very difficult to determine for a specific quantum system, unless that system is very low dimensional or possesses a very high degree of symmetry. Nonetheless, in some special cases the limit can be computed; see for example \cite{khaneja2001, khaneja2002, boscain2002, carlini2006, khanejaHeitmann2007, hegerfeldt2013, dalessandro2020}. This difficulty means that work has concentrated on finding lower bounds for the speed limit rather than exact results. Various bounds have been obtained for closed, finite dimensional systems as well as for open systems  \cite{margolus1998, giovannetti2003, kosinski2006, jones2010, russell2014, kallush2014, lam2021, delcampo2013, deffner2013, taddei2013, liu2015, pires2016, lee2020, shao2020,burgarth2022}. While these bounds are not tight, they can provide information on how the speed limit is likely to scale with regard to quantities of interest, such as system dimension or total energy. It is notable that many of these approaches make use of energy uncertainty of the system, applying the original results of Mandelstam and Tamm \cite{mandelstam1945}, as well as the more modern interpretation of Margolus and Levitin \cite{margolus1998}.

Given this background, we can state a generic quantum control problem and investigate its speed limit as follows. We consider a Hamiltonian given by
\begin{equation}\label{eqControlProblem}
H = H_d + \sum_{j=1}^m f_j(t) H_j
\end{equation}
where $H_d$ and $H_j$ are time-independent Hamiltonians acting on a finite-dimensional Hilbert space, and the $f_j(t)$ are a set of real, time-dependent, scalar functions. $H_d$ is called the {\emph{drift Hamiltonian}}, and is always present. The $H_j$ are the {\emph{control Hamiltonians}}, and we assume that we have arbitrary control over the $f_j(t)$, as even in this case the quantum speed limits are very difficult to determine.

The system evolves according to the Schr{\"{o}}dinger equation 
\begin{equation}
i \frac{d}{dt} U(t) = \bigg( H_d + \sum_{j=1}^m f_j(t) H_j \bigg) U(t), \,\,\,\,\, U(0) = {\mathbb{1}},
\label{eqSchroedingerEquation}
\end{equation}
where $U(t)$ is the unitary time-evolution operator. In an $n$-dimensional system $U(t)$ can be represented as a unitary $n\times n$ matrix. Further, as unitary operators are physically indistinguishable up to a phase, we can choose to remove this excessive phase degree of freedom by demanding that $U(t)$ have unit determinant, making it a special unitary matrix. This is accomplished by choosing the drift and control Hamiltonians to be traceless, and we will assume this is the case throughout this paper.

The system is called {\emph{controllable}} if it is possible to find control functions $f_j(t)$ such that, given enough time, we can achieve any possible unitary (up to a phase), or equivalently, if we can generate all possible members of the Lie group SU$(n)$. There is a beautiful Lie-algebraic result that states that this is the case if and only if \cite{jurdjevic1972, dalessandro2008} the dynamical Lie algebra $\{ iH_d, iH_1, iH_2, \ldots , iH_m \}_{LA}$ has dimension $n^2-1$, i.e.\ the dynamical Lie algebra generated by the control Hamiltonians and drift Hamiltonian is the Lie algebra ${\mathfrak{su}(n)}$.

The next natural question is, if a quantum system is controllable, how long will it take to produce a specific unitary in the worst case? Or, equivalently \cite{jurdjevic1996,elliot2009} in the case of compact groups such as $\text{SU}(n)$, since the system is controllable, what is the minimum time by which we can guarantee we can produce all possible unitaries? This is what we will refer to as the {\emph{quantum speed limit}} in this paper.

We note that some authors make a distinction between quantum control systems which are fully controllable only in the presence of a drift term (i.e.\ removing the drift Hamiltonian would cause the system to no longer be fully controllable) from those systems for which this is not the case. Systems of the latter type are known as strongly controllable \cite{arenz2017}, and are fully controllable with control Hamiltonians alone regardless of the presence or absence of any drift term. Due to our assumption that the control strengths $f_j(t)$ can be arbitrarily large, strongly controllable systems can reach any unitary in an arbitrarily short amount of time, rendering the concept of a speed limit irrelevant. For that reason we consider only systems of the first type, where the drift is required to ensure the system is controllable.

In this paper, our goal is to derive lower bounds on the speed limits of controllable quantum systems that are as general as possible. We do not restrict the system to a specific number of dimensions, or demand it describes a set of qubits, or require the drift Hamiltonian to be of a specific form, as is common in other speed limit calculations. We require no knowledge of the quantum energy uncertainty of the system. 
We will require only that the control subgroup is topologically closed, where the control subgroup is the set of all unitaries that can be reached by application of the control Hamiltonians alone, and will thus form a subgroup of $\text{SU}(n)$. 
However the resulting speed limits can be hard to analytically compute explicitly as they require determining the diameter of rather abstract manifolds, so we examine in more detail cases where the manifolds are symmetric spaces \cite{gallier}, which can arise, for example, if the Lie algebra associated with the control subgroup forms a Cartan decomposition \cite{dalessandro2008} of the full dynamical Lie algebra.

This will allow us to derive explicit analytic lower bounds for the quantum speed limit for a number of control schemes corresponding to cases where the control group is one of $\text{SO}(n)$, $\text{Sp}(n/2)$, or $\text{S}(\text{U}(p) \times \text{U}(q))$ with $p+q=n$, and investigate when this bound will be tight. We also consider the case where the number of control Hamiltonians is not enough to span the full Lie algebra corresponding to these groups, and give the minimum number of control Hamiltonians required to generate the algebra. Due to the fact that many control problems will not have enough controls to generate these groups, we also derive a bound for the common general case where there is a only a single control Hamiltonian. In all cases, our results are completely general and valid for arbitrary dimension. Finally, in order to test our analytic results, we carry out an exploration of quantum speed limits for a variety of low dimensional systems using numerical simulations. This not only provides a check on our results, but allows an investigation of the efficiency of numerical optimal control algorithms for bilinear systems.

The structure of the paper is as follows. We begin in Section~\ref{secProblemFormulation} by formulating the quantum speed limit problem in terms of Lie algebras and Lie groups, and introduce concepts we will require such as cosets, quotient spaces and adjoint orbits, as well as laying out our basic approach. We introduce the idea that the problem can be treated as movement on a manifold, with the movement direction and speed given by the drift Hamiltonian. Since the mathematical machinery will not be familiar to some readers, we provide illustrative examples.

In Section~\ref{secQSLDiamAndDrift} we explain how one can obtain a speed limit by determining the diameter of a manifold (i.e.\ the two points furthest apart) and dividing by the speed at which the system moves on the manifold. We describe the conditions on the manifold required for this to work, and give a way of computing the speed of movement from the system's drift Hamiltonian. We establish that symmetric spaces provide manifolds meeting the criteria, give their diameters, and use them to compute explicit expressions for the lower bound on the quantum speed limit. 

Section~\ref{secBoundTightnessDimensionCounting} examines when the lower bound developed in the previous section is actually tight. It develops a criterion based on the dimension of the adjoint orbit and commutation relations between the drift Hamiltonian and the matrix representation of the Lie algebra corresponding to the controls.

As this criterion is sufficient but not necessary, in Section~\ref{secQSLWithCartanControls} we investigate what else can be said about the tightness of the bound if the controls arise from a Cartan decomposition. This allows understanding the control problem in terms of root systems, and we illustrate the results by considering the case where the control group is $\text{SO}(n)$.

In Section~\ref{secNumericalTests} we treat the problem of finding the quantum speed limit numerically, and compare the simulations to our analytic results. This allows both a test of our bounds as well as an examination of how well standard optimization techniques used in quantum control work.

Finally, in Section~\ref{secQSLWithoutFullCartan} we consider the case where we have only a limited set of controls so that we do not have a symmetric space, and derive a bound on the speed limit for the common case where the system has only a single control Hamiltonian.

\section{Problem formulation in terms of Lie groups and algebras}
\label{secProblemFormulation}

The calculation of quantum speed limits is often approached using Lie group-theoretic techniques. We will also make use of these mathematical structures, so we briefly provide the relevant background here. Good explanations of this material can be found in, for example, \cite{dalessandro2008, gallier, hall2015}.

In what follows, we will denote groups with a Roman upper case letter, e.g.\ $G = \text{SU}(n)$, and algebras with a Fraktur font (e.g.\ $\mathfrak{g} = \mathfrak{su}(n)$).

Let $\mathfrak{g}$ be the full Lie algebra generated by the drift Hamiltonian and the control Hamiltonians, i.e.\ ${\mathfrak{g}} = \{iH_d, iH_1, iH_2, \ldots , iH_m \}_{LA}$, and let the Lie algebra generated by the control Hamiltonians alone be given by ${\mathfrak{k}} = \{iH_1, iH_2, \ldots , iH_m \}_{LA}$. 
Clearly ${\mathfrak{k}}$ is a subalgebra of ${\mathfrak{g}}$. The system is said to be controllable if ${\mathfrak{g}} = {\mathfrak{su}}(n)$. 

We denote the control group, i.e.\ the group of unitaries generated by exponentiating ${\mathfrak{k}}$, by $K$, and the dynamical Lie group generated by ${\mathfrak{g}}$ with $G$. 
Clearly $K \subseteq G$ is a subgroup, and $G \subseteq \text{SU}(n)$ with equality if the system is controllable.

At any given time, the system evolves according to (\ref{eqSchroedingerEquation}). Since the control amplitudes $f_j(t)$ can be arbitrarily large, we can  generate any unitary $U \in K$ in an arbitrarily short time to arbitrarily good precision (see \cite{dalessandro2008} for a rigorous justification of this point). 
Now, suppose our control problem is to produce a unitary $U_\text{target}$ that moves us between the two unitaries $U_1$ and $U_2$, i.e.\ $U_2 = U_{\text{target}} U_1$. 

Since we can move between elements of $K$ arbitrarily quickly, all elements of $K$ are equivalent, meaning if we apply any controls after we have generated the specific unitary $U_{\text{target}}$, all resulting unitaries $K U_{\text{target}}$ are equivalent in terms of how quickly we can generate them. Because of this, we can view our control problem as actually asking how to move between the right cosets $K U_1$ and $K U_2$, where the right coset is $K U = \{ k U | k\in K\}$. Furthermore, as the system evolves in time, the unitary at any point in time, given by (\ref{eqSchroedingerEquation}), is equivalent to any other element in its coset, because it can be moved within the coset arbitrarily quickly.

Alternatively, one could define equivalence in terms of \emph{left} cosets, where now we consider how to move between the left cosets $U_1 K$ and $U_2 K$. Again, these cosets are equivalent in terms of the minimum time it takes to use controls to move between them, but now the controls are being applied before the unitary rather than after. 

From a quantum control perspective, the first description is more intuitive. That is, if we are given a unitary, another unitary is equivalent if we can move to it arbitrarily quickly. This corresponds to the right coset picture, since applying controls to a given unitary corresponds to left multiplication by the controls. Consequently this is the approach we will take for this paper.


These arguments show that the relevant elements of the control problem are the cosets, and an effective way of formulating the problem is to ``divide out'' the degree of freedom associated with each coset. To make this  rigorous one defines the quotient space $G/K$ as the set of right cosets $K g, \,\, \forall g\in G$. 
We denote each coset by $[ g ] = Kg$ with $g \in G$, since the element $g$ indexes the coset. The cosets can also be seen as the orbits of the natural left action of $K$ on $G$, and the space of orbits is $G/K$.

If $K$ is a normal subgroup of $G$, then $G/K$ is itself a Lie group \cite{hall2015, gallier}. 
However, even if this is not the case, provided $G$ is a Lie group, and $K$ is a {\emph{closed}} subgroup (in the topological sense) then $G/K$ is a differentiable manifold \cite{hall2015} that is also a (right) homogeneous space meaning that it carries a (right) transitive $G$-action which is given by $[g']\cdot g=[g'g]$.
Specifically, $G/K$ can be given the structure of a smooth manifold with dimension $\text{dim}(G/K) = \text{dim}\, G - \text{dim}\, K$. Movement within a coset does not result in movement in $G/K$, but movement between cosets does. Movement within a coset is produced by the control Hamiltonians, and movement between the cosets requires the drift Hamiltonian.


As the system evolves via (\ref{eqSchroedingerEquation}) it traces out a continuous path in $G/K$ space, and the quantum speed limit is governed by how fast we can move between the two points corresponding to $U_1$ and $U_2$. Clearly we cannot move arbitrarily in $G/K$. 
Our movement on $G/K$ is determined by the drift Hamiltonian, with the direction of the movement determined by where we are within a coset at any given time, allowing us some degree of steering.

\hide{
To make this concept more precise, we need some concepts from differential geometry.
We pick an Ad-invariant inner product on $\mathfrak g$ (e.g.\ the Hilbert-Schmidt inner product).
Setting $\mathfrak p = \mathfrak g^\perp$ so that $\mathfrak g = \mathfrak k \oplus \mathfrak p$ we obtain a decomposition which automatically satisfies
\begin{eqnarray}
{[}{\mathfrak{k}}, {\mathfrak{p}}{]} & \subseteq & {\mathfrak{p}} \label{eqReductiveHP}\\
{[}{\mathfrak{k}}, {\mathfrak{k}}{]} & \subseteq & {\mathfrak{k}}
\end{eqnarray}
where the first equality is due to Ad-invariance and the second inequality follows from $\mathfrak k$ being the Lie algebra of the subgroup $K$.
The $\mathrm{Ad}$-invariant inner product on $\mathfrak g$ induces a bi-invariant Riemannian geometry on $G$ which in turn induces a $G$-invariant Riemannian geometry on $G/K$ (see the next section for details).
This equips the manifold $G/K$ with the structure of a naturally reductive space (are more restricted variety of homogeneous spaces).
}

In particular, we have the following: If $G$ is a compact and connected Lie group (e.g. $\text{SU}(n))$, and $K$ is a closed Lie subgroup of $G$, with associated Lie algebras ${\mathfrak{g}}$ and ${\mathfrak{k}}$, then we can decompose the Lie algebra ${\mathfrak{g}}$ as ${\mathfrak{g}} = {\mathfrak{p}} \oplus {\mathfrak{k}}$ with
\begin{eqnarray}
{[}{\mathfrak{k}}, {\mathfrak{p}}{]} & \subseteq & {\mathfrak{p}} \label{eqReductiveHP}\\
{[}{\mathfrak{k}}, {\mathfrak{k}}{]} & \subseteq & {\mathfrak{k}}
\end{eqnarray}
where ${\mathfrak{p}} = {\mathfrak{k}}^\perp$ with respect to an Ad-invariant inner product on $\mathfrak g$ \cite{gallier}, e.g.\ the Hilbert-Schmidt inner product. Note that while ${\mathfrak{k}}$ is a Lie algebra, ${\mathfrak{p}}$ is in general not closed under the Lie bracket.
The $\mathrm{Ad}$-invariant inner product on $\mathfrak g$ induces a bi-invariant Riemannian geometry on $G$ which in turn induces a $G$-invariant Riemannian geometry on $G/K$ (see the next section for details).
This equips the manifold $G/K$ with the structure of a so-called {\emph{reductive space}}, which is a more restricted variety of a homogeneous space.

Any evolution purely under the action of the controls, without the drift, will produce motion only within a coset. Without loss of generality, we can assume $iH_d \in {\mathfrak{p}}$, since any contribution that lies in ${\mathfrak{k}}$ can be removed by application of the controls. 
Since $\mathfrak{p}$ is orthogonal to $\mathfrak k$, this means that any evolution under the drift alone moves purely in $G/K$, with no movement within a coset. 
Specifically, for a reductive space, the inner product lets us identify the tangent $T_o(G/K)$ at the origin $o=[\mathbb1]$ with $\mathfrak{p}$.

To show how the action of the control steers the direction of motion in $G/K$, we need the concept of the adjoint orbit. The adjoint orbit of $A\in{\mathfrak{g}}$ is given by
\begin{equation}
{\cal{O}}(A) = \{ k^{-1} A k \,| \, k\in K \}.
\end{equation}
By Eq.~\eqref{eqReductiveHP}, we have $\mathcal O(A)\subset \mathfrak p$ for $A\in\mathfrak p$.
We can see how this steers the evolution in $G/K$ space as follows \cite{khaneja2001}: Take elements $k_1, k_2 \in K$ that belong to the coset containing the identity, and consider where they move under the action of the drift after a short time $\Delta t$. We obtain
\begin{equation}
k_1 \rightarrow e^{-i H_d \, \Delta t} k_1 = k_1 e^{-i k_1^{-1} H_d k_1 \Delta t}
\end{equation}
showing that after the evolution it is now a member of the coset $[e^{-i k_1^{-1} H_d k_1 \Delta t}]$. Similarly, $k_2$ moves to a coset $[e^{-i k_2^{-1} H_d k_2 \Delta t}]$. 
Since we can choose to be anywhere in a coset arbitrarily quickly due to the action of the controls, we see that the adjoint orbit represents the directions we are able to move from the origin of $G/K$.

This mathematical machinery can be somewhat opaque, so we present a simple example that illustrates these concepts. We consider computing the quantum speed limit of a controllable quantum system in a two dimensional Hilbert space, i.e.\ the group associated with the unitary evolution operator is $\text{SU}(2)$. 
This is one of the few cases where the speed limit is explicitly known.

We take our Hamiltonian to be
\begin{equation}
H = \sigma_z + f(t) \sigma_x
\label{eqHSU2ControlProblem}
\end{equation}
and the Schr{\"{o}}dinger equation is given by
\begin{equation}
-i\frac d{dt}{U}(t) = (\sigma_z + f(t) \sigma_x)U(t), \quad U(0)=\mathbb1.
\end{equation}
The Lie algebra associated with the single control is just $\text{span} \{ i \sigma_x \}$, while the full dynamical Lie algebra associated with the drift and controls is $\text{span} \{ i\sigma_x, i\sigma_y, i\sigma_z\}$. 
Since this algebra is three dimensional, and this matches $n^2-1$ where $n$ is the Hilbert space dimension, the system is controllable. 
Our Lie algebra decomposition is ${\mathfrak{g}} = {\mathfrak{p}} \oplus {\mathfrak{k}}$ with ${\mathfrak{k}} = \text{span} \{ i \sigma_x \}$ and ${\mathfrak{p}} = \text{span} \{i\sigma_y, i\sigma_z \} $. 
We have ${\mathfrak{g}} = {\mathfrak{su}}(2)$, ${\mathfrak{k}} = {\mathfrak{u}}(1)  $, $G=\text{SU(2)}$ and $K=\text{U}(1)$. 
The manifold corresponding to the quotient space $G/K$ can in general be quite complicated, but in this case it is particularly simple; the manifold $G/K = \text{SU(2)}/\text{U(1)}$ is isomorphic to the two-sphere $S^2$.

Since the control algebra is one-dimensional, the control group subgroup $K$ generated by $\mathfrak{k}$ can be parameterized by a single parameter $\alpha$ as $e^{i \alpha \sigma_x}, \,\, \alpha \in [0, 2\pi]$, and the adjoint orbit is given by the set
\begin{align}
{\cal{O}}(iH_d) &=  \{ e^{-i \alpha \sigma_x} i \sigma_z e^{i \alpha \sigma_x} |\,\alpha\in[0,2\pi]\}\nonumber \\
&=  \{ i \cos (2 \alpha)\, \sigma_z + i \sin (2\alpha)\,\sigma_y |\,\alpha\in[0,2\pi]\}
\label{eqSU2SimpleExampleOrbit}.
\end{align}
$S^2$ is two-dimensional, and the tangent space at the origin is defined by $\text{span} \{ i \sigma_y, i\sigma_z \} = {\mathfrak{p}}$. Since Eq.~(\ref{eqSU2SimpleExampleOrbit}) allows any direction in the tangent space by suitable choice of $\alpha$, we can move in any direction in $G/K$ we wish. 
As we will show in later sections, the speed of movement in $G/K$ is constant and determined purely by the drift Hamiltonian. This means that the speed limit is achieved by moving on a great circle geodesic between two antipodal points, as this yields the maximum possible evolution time between any two unitaries for the system.

 The concepts of speeds and distances on the $G/K$ manifold are determined by the Riemannian metric on $G/K$ which depends on the inner product chosen on ${\mathfrak{g}}$. 
 As will be shown later, if we choose the Killing form for the inner product, then for this particular example the speed of movement is $2\sqrt{2}$, and the distance between two antipodal points is $\sqrt{2} \pi$, giving the time for the quantum speed limit as $t=\pi/2$, which agrees with the standard result \cite{dalessandro2008}.

We note that this is an unusual way to look at this problem. The normal approach is to apply the maxim ``algebra is easier than geometry'', and use Lie algebra, Lie groups and results such as the maximal tori theorem, rather than considering geodesics on a manifold. 
Nonetheless, the idea of obtaining a speed limit by dividing the diameter of the $G/K$ manifold by the drift velocity will prove extremely useful. In the case where the adjoint orbit allows us to move on a geodesic connecting the two points furthest apart on the manifold, we can obtain an exact speed limit, and if it does not allow movement on such a geodesic, such a method will still provide a lower bound.

\section{Quantum speed limits from manifold diameter and drift velocity}
\label{secQSLDiamAndDrift}

As discussed in the previous Section, in order to obtain speed limits from the structure of the $G/K$ manifold, we need some way of assigning distances to the space. This involves bridging the two descriptions of the problem: The control and drift Hamiltonians defining the system are described by the Lie algebra, while the unitaries corresponding to the system evolution are described by the Lie group and associated manifold.

To see the issue, consider the group $\text{SU}(2)$. The associated manifold is the three-sphere, which describes the topology, but there is no metric associated with it (yet) --- there is no concept of the size of its diameter, for example. 
The way the metric is imposed is to define an inner product on the Lie algebra which is then pushed around the group to define an inner metric on all tangent spaces. 
\hide{While there are arbitrarily many inner products that can be chosen, for semisimple compact groups (e.g.\ $\text{SU}(n)$) there is one canonical choice, the Killing inner product which is also Ad-invariant. 
It takes two elements $X,Y$ of a Lie algebra and is given by 
\begin{equation}
\langle X,Y\rangle_K = -\text{Tr}[\text{ad}_X \circ \text{ad}_Y]
\end{equation}
where $\text{ad}_X = [X,\cdot\,]$ is the adjoint representation of $X$ acting on $\mathfrak g$. It has a particularly simple form for $\mathfrak{su}(n)$ given by
\begin{equation}
    \langle X,Y\rangle_K = -2n \, \text{Tr}[XY] \label{eqKillingFormSUN}
\end{equation}
where $X,Y\in\mathfrak{su}(n)$. 
}
For the inner product on the Lie algebra $\mathfrak g$ we will take
\begin{equation}\label{eqKillingFormSUN}
    \langle X,Y \rangle_K = -2n \, \mathrm{Tr}[XY], \quad X,Y\in\mathfrak g.
\end{equation}
This inner product is Ad-invariant since the group $G$ consists of unitary operators.
In the controllable case, i.e., $\mathfrak g = \mathfrak{su}(n)$, this is the Killing form.
We now obtain the inner product at the tangent space of a general element $g\in G$ from
\begin{equation}\label{eqRiemannianMetric}
    \langle X,Y\rangle_g = \langle g^{-1}X,g^{-1}Y\rangle_{K} = \langle X g^{-1},Yg^{-1}\rangle_K
\end{equation}
where $X,Y\in T_gG$ are tangent vectors at $g$. The second equality holds by Ad-invariance of the inner product on $\mathfrak g$.
This equips $G$ with a bi-invariant Riemannian geometry (meaning that both left and right multiplication act isometrically).
For such groups the geodesics through an element $g$ are precisely the curves of the form $t\mapsto ge^{vt}$, where $v \in {\mathfrak{g}}$ \cite{milnor1969} (Lemma 21.2).

The quotient space $G/K$ inherits a $G$-invariant Riemannian geometry from $G$: At the origin $o$ the inner product $\langle X,Y\rangle_o$ is defined as $\langle X,Y\rangle_K$ using that $T_oG/K\cong \mathfrak p \subset \mathfrak g$. 
This is extended to arbitrary points $[g]$ by the (differential of the) $G$-action just as in \eqref{eqRiemannianMetric}: $\langle X,Y \rangle_{[g]}= \langle X\cdot g^{-1}, Y\cdot g^{-1} \rangle_o$ (this is indeed well-defined, i.e.\ independent of the choice of $g$ within the coset).
In particular, the resulting Riemannian metric is automatically $G$-invariant meaning that $G$ acts isometrically on $G/K$.
It now holds by construction that the natural projection $\pi : G \to G/K$ is a Riemannian submersion \cite{gallier} meaning 
it induces an isometry between $(\ker d\pi|_g)^\perp$ and $T_{\pi(g)}(G/K)$ for all $g$. 
Since $\ker(d\pi|_{\mathbb1})=\mathfrak k$, this just follows from $T_o(G/K) \cong \mathfrak p$ and our definition of the metric (in general, we have $\ker d\pi|_g = g^{-1}\mathfrak k g$ and hence $T_{[g]}G/K \cong g^{-1}\mathfrak pg$). 
The notation $d\pi|_g$ means that we take the differential of $\pi$ at the point $g$ which is a linear map $T_gG\to T_{\pi(g)}G/K$. 

The crucial point for us is the following: From $\pi$ being a Riemannian submersion it follows that geodesics in $G/K$ running through a coset $x=[g]$ are precisely curves of the form $[g\exp(ut)]=x\cdot\exp(ut)$ with $u\in g^{-1}\mathfrak pg$ and that they have the same length as their corresponding lifts of $G$ \cite{gallier} (Proposition 18.8).

Let us summarize the relevant structure: We have a quantum control problem with dynamical Lie algebra $\mathfrak g$ and control algebra ${\mathfrak{k}}$, associated Lie groups $G=e^{\mathfrak{g}}$ and $K = e^{\mathfrak{k}}$, and $K$ is a closed subgroup of $G$. 
We use the Killing form as an inner product on ${\mathfrak{g}}$ and take the decomposition ${\mathfrak{g}} = {\mathfrak{p}} \oplus {\mathfrak{k}}$ with ${\mathfrak{p}} = {\mathfrak{k}}^\perp$. 
We can always ensure $iH_d \in {\mathfrak{p}}$ by removing any part not in ${\mathfrak{p}}$ via the controls. We know $G/K$ is a reductive space and we know precisely which form the geodesics on $G/K$ have.


We now compute the speed at which the system moves through $G/K$ as it evolves. 
We know the possible directions of travel at the origin are given by the adjoint orbit of the drift, $ k^\dagger i H_d k\in{\mathcal{O}}(iH_d)$ with $k \in K$, so that in a time $dt$ we move to a coset of $\exp (i k^\dagger H_d k \, dt)$. 
To determine the distance $ds$ this corresponds to in $G/K$ we use the metric on $G/K$ and because we have a Riemannian submersion we can employ (\ref{eqKillingFormSUN}) to obtain
\begin{eqnarray}
ds &=& \sqrt{ \langle i k^{-1} H_d k \, dt, i k^{-1} H_d k \, dt\rangle_K} \nonumber \\
&=& dt \sqrt{2n \, \text{Tr}(H_d^2)}
\end{eqnarray}
where we have used the fact that the Killing form is Ad-invariant.
Using the $G$-invariance of the metric on $G/K$, the same argument shows this result also holds at other points $x\ne o \in G/K$.
This means the speed at which the system moves in $G/K$ is constant and is given by
\begin{equation}
v = \sqrt{2n \, \text{Tr} \left( H_d^2 \right)}.
\label{eqSpeedInGOverK}
\end{equation}

Now that we know the form a geodesic in $G/K$ must take, and speed with which a quantum system moves along it, the task is to find the diameter of the $G/K$ space, that is, the furthest distance possible pairs of points can have. 
Given the fact that motion in $G/K$ is at constant speed, this will give us a lower bound on the quantum speed limit, that is, the time taken to produce the most difficult unitary.
This proves the following:
\begin{theorem}
    Let $G$ be the dynamical Lie group of the control problem Eq.~\eqref{eqControlProblem} and assume that the subgroup $K\subset G$ generated by the controls alone is closed.
    Let $T_{\mathrm{QSL}}$ be the minimum time in which all unitaries of $G$ can be reached. Then 
    \begin{equation}\label{eqQSL}
        T_{\mathrm{QSL}} \ge \frac{\mathrm{diam}(G/K)}{\sqrt{2n \, \mathrm{Tr}[H_d^2]}}.
    \end{equation}
\end{theorem}

The practical usefulness of this result, of course, relies on an explicit computation of the diameter (or at least a lower bound).
The diameter of the Riemannian manifold $G/K$ is
\begin{align}
    \text{diam}(M) = \sup \{ d(x,y): x,y \in G/K\}
\end{align}
where $d(x,y)$, the Riemannian distance between $x$ and $y$, is the infimum over the lengths of curves connecting these points as measured by the metric.
Since $G/K$ is homogeneous the definition is equivalent to $\mathrm{diam}(M)=\sup_{x\in G/K} d(x,o)$.

That Eq.~\eqref{eqQSL} is only a lower bound in general is due to the restricted movement on $G/K$: The possible directions are given by the adjoint orbit $\mathcal O (iH_d)$.
If the adjoint orbit does not allow for the needed directions, the time taken to generate some unitaries will be longer than the lower bound given in Eqs.~(\ref{eqSpeedLimitSOEven}) -- (\ref{eqSpeedLimitUpUq}).

Finding the diameter of the homogeneous space $G/K$ is in general difficult.
However, the diameter of all symmetric spaces arising from classical compact groups has been calculated by Yang \cite{yang2007}. (We note there appears to be an error in Yang's paper; the results given for the diameters of $\text{SU}(2n)/\text{Sp}(n)$ should be divided by $\sqrt{2}$). 
If we consider only symmetric spaces arising from quotient groups of the form $G/K$ where $G = \text{SU}(n)$, there are only three possibilities, which we list in Table~\ref{tabDiameters}. Note that the group $\text{Sp}(n)$ refers to the compact symplectic group, and we have chosen to use the Killing form as the inner product on the Lie algebra $\mathfrak{g}$ to obtain a metric on $G/K$.


\renewcommand{\arraystretch}{2.0}
\begin{table}[ht!]
\begin{center}
\begin{tabular}{  c c  } 
$G/K$ & $\text{diam}(G/K)$   \\
\hline \\[-4ex]
$\text{SU}(n)/\text{SO}(n)$ & $ \begin{dcases}
     \,\, \frac{\sqrt{2}}{2} \pi n & \text{if $n$ even} \\
     \,\, \frac{\sqrt{2}}{2} \pi (n^2 -1)^{1/2} & \text{if $n$ odd}
    \end{dcases} $    \\  [25pt]
$\text{SU}(2n)/\text{Sp}(n)$ &  $\begin{dcases}
     \,\,  \pi n & \text{if $n$ even} \\
     \,\, \pi (n^2 -1)^{1/2} & \text{if $n$ odd}
    \end{dcases} $    \\ 
$\text{SU}(p+q)/\text{S}(\text{U}(p) \times \text{U}(q)) $ & $\pi(p+q)^{1/2} p^{1/2}  \,\,\,\, p\leq q $ \\
\hline
\end{tabular}
\end{center}
\caption{The diameter of various compact symmetric spaces arising from the quotient $G/K$, when using the Killing inner product on $\mathfrak{g}$ in order to obtain a Riemannian metric on $G/K$.}
\label{tabDiameters}
\end{table}
\renewcommand{\arraystretch}{1.0}

Consequently if the Lie group $K$ generated by the controls is one of $\text{SO}(n)$, $\text{Sp}(n)$, or $\text{S}(\text{U}(p) \times \text{U}(q))$ (the matrices of unit determinant in $\mathrm U(p)\times \mathrm U(q)$), we obtain the following quantum speed limits: 
\begin{numcases}
{\text{SO}(n): \,\,\,\, T_{\text{QSL}}  \ge } 
     \,\, \frac{ \sqrt{n} \, \pi}{2\sqrt{ \, \text{Tr} (H_d^2)}}  & \text{if $n$ even} \label{eqSpeedLimitSOEven} \\
     \,\, \frac{\pi (n^2 -1)^{1/2}}{2\sqrt{n \, \text{Tr} (H_d^2)}} & \text{if $n$ odd} \label{eqSpeedLimitSOOdd}
\end{numcases}

\begin{numcases}
{\text{Sp}(n): \,\,\,\, T_{\text{QSL}}  \ge } 
     \,\, \frac{ \sqrt{n} \, \pi}{\sqrt{ 2 \, \text{Tr} (H_d^2)}}  & \text{if $n$ even} \label{eqSpeedLimitSpEven} \\
     \,\, \frac{\pi (n^2 -1)^{1/2}}{\sqrt{2 n \, \text{Tr} (H_d^2)}} & \text{if $n$ odd} \label{eqSpeedLimitSpOdd}
\end{numcases}

\begin{equation}
\text{S}(\text{U}(p) \times \text{U}(q)): \,\,\,\, T_{\text{QSL}}  \ge
     \,\, \frac{ \sqrt{p} \, \pi}{\sqrt{ 2 \, \text{Tr} (H_d^2)}} \,\,\,\, p\leq q   \label{eqSpeedLimitUpUq} 
\end{equation}

The result for the case where the control group is $\text{Sp}(n)$ is particularly interesting. It is known that this control group provides complete {\emph{state}} controllability even in the absence of a drift Hamiltonian \cite{dalessandro2008}. As we have assumed arbitrarily strong controls, this means that one can find controls that move from any state to any other state arbitrarily quickly. That is, the speed limit for state control in this case is zero. The emergence of a finite speed limit as given by (\ref{eqSpeedLimitSpEven}) and (\ref{eqSpeedLimitSpOdd}) highlights the difference between unitary control and state control.

It is also worth noting the appearance of explicit dependence of the Hilbert space dimension in these bounds, as existing speed limits in the literature are usually not able to include this factor.

\section{Bound tightness in terms of dimension counting}
\label{secBoundTightnessDimensionCounting}

Let us discuss the tightness of our speed limit bounds from the perspective of the dimensions of the control group. Our bound was obtained from the observations that the speed of movement in $G/K$ is constant and that the largest distance between two points (the diameter) is finite.
While the existence of a length minimizing geodesic connecting the origin $o$ with any other point $x\in G/K$ is guaranteed (by the Hopf-Rinow theorem), it is not clear that such a geodesic is available by choice of suitable controls.

Denote by $D$ the set of points maximizing the distance from the origin, i.e.\ the points $x\in G/K$ with $d(o,x) = \mathrm{diam}(G/K)$ where $d$ dentotes the Riemannian length on $G/K$. 
As both inversion and the $K$-action are isometries that fix the origin, we know that they also leave $D$ invariant, i.e.\ $D^{-1}=D$ and $D\cdot k =D$ for all $k\in K$.
For the bounds to be tight, it is necessary and sufficient that \emph{for each} $x\in D$, there is a minimal geodesic connecting the origin $o$ with $x$ which is of the form $[\exp(vt)]$ with $v\in \mathcal O(iH_d)$.
This trivially holds if $\mathcal O(iH_d)$ is equal to the sphere $S=\partial B_r(0)$ in $\mathfrak p$ of radius $r=\sqrt{\langle H_d,H_d\rangle_K}$ (note that all directions in the adjoint orbit have the same length by Ad-invariance).
The adjoint orbit itself is a closed manifold which is a subset of $S$.
In the case that the dimension of $\mathcal O(iH_d)$ is maximal (i.e.\ equal to $\dim \mathfrak p - 1$), it follows that $\mathcal O(iH_d)$ is equal to $S$ and thus contains every direction in $\mathfrak p$.

\renewcommand{\arraystretch}{2.0}
\begin{table}[ht!]
\begin{center}
\begin{tabular}{ c c c c} 
$G/K$ & $d_p$ &  $d_k$  & $d_k - d_p  + 1 $ \\
\hline \\[-4ex]
$\text{SU}(n)/\text{SO}(n)$           & $\frac{1}{2} (n^2\!+\!n\!-\!2)$   & $\frac{n}{2} (n-1)$ & $2-n$ \\
$\text{SU}(n)/\text{Sp}(\frac{n}{2})$ & $\frac{1}{2} (n^2\!-\!n\!-\!2)$   & $\frac{n}{2} (n+1)$ &  $2+n$ \\ 
$\text{SU}(p\!+\!q)/\text{S}(\text{U}(p)\!\times\! \text{U}(q)) $ & $2pq $  &  $ p^2\! +\! q^2\!-\!1$  &  $(p\!-\!q)^2$ \\
\hline
\end{tabular}
\end{center}
\caption{The dimensions of the Lie algebras associated with the three symmetric spaces associated with $\text{SU}(n)$. $d_k= \text{dim}({\mathfrak{k}})$ is the dimension of the control algebra and $d_p= \text{dim}({\mathfrak{p}})$ is the dimension of the symmetric space $G/K$. If $\dim(\{A\in \mathfrak k\,|\,[H_d,A]=0\}) = 1+ \dim \mathfrak k - \dim \mathfrak p$, the adjoint orbit from the controls is guaranteed to have enough degrees of freedom to choose any single parameter geodesic from the origin to a point corresponding to the diameter of the space.}
\label{tabDimensions0}
\end{table}
\renewcommand{\arraystretch}{1.0}

The dimension of the adjoint orbit is
\begin{equation}\label{eqDimensionAdjointOrbit}
    \dim\mathcal O(iH_d) =\dim\mathfrak k -\dim(\{A\in \mathfrak k\,|\,[H_d,A]=0\})
\end{equation}
because $T_A \mathcal O(A) \cong \mathfrak p / \ker [A,\cdot\,]$.
This means that the bound is tight if we have equality in 
\begin{equation}\label{eqDimensionCountingCrit}
    \dim(\{A\in \mathfrak k\,|\,[H_d,A]=0\}) \ge 1+ \dim \mathfrak k - \dim \mathfrak p.
\end{equation}
This inequality always holds and equality is equivalent to the ability to move into every possible direction in $G/K$.

We stress that this is a \emph{sufficient} condition, but not a necessary one.  Even if the adjoint orbit does not have enough directions to access all dimensions of ${\mathfrak{p}}$, that does not rule out the possibility that, for a specific drift Hamiltonian, a single-parameter geodesic from the origin to the locus corresponding to the diameter with an initial direction lying in the adjoint orbit does not exist.

Table~\ref{tabDimensions0} lists the relevant dimensions for $\mathfrak{k}$ and $\mathfrak{p}$ for the symmetric spaces we are considering, as well the quantity corresponding to the right hand side of (\ref{eqDimensionCountingCrit}). For the symmetric spaces $\text{SU}(n)/\text{Sp}(\frac{n}{2})$ and  $\text{SU}(p + q)/\text{S}(\text{U}(p) \times \text{U}(q)) $ the number degrees of freedom in the control group exceeds that of the quotient space, so naive dimension counting arguments suggest the bound is likely to be tight, although one must test for equality in Eq.~(\ref{eqDimensionCountingCrit}) to be sure.

However, it is clear that for the case $\text{SU}(n)/\text{SO}(n)$ with $n>2$ it is never possible to achieve equality in (\ref{eqDimensionCountingCrit}) as the dimension of a space can never be less than zero. Nonetheless, as we will see in our numerical tests of the speed limit in Section~\ref{secNumericalTests}, for some drift Hamiltonians the bounds (\ref{eqSpeedLimitSOEven}) and (\ref{eqSpeedLimitSOOdd}) are still tight. To investigate this in more detail, we consider case where the control algebra is ${\mathfrak{k}} = \mathfrak{so}(n)$. We wish to determine the size of $\dim(\{A\in {\mathfrak{k}} \,|\,[H_d,A]=0) \}$. To begin, we note that any drift $H_d$ can be moved into the Cartan subalgebra by some controls. This subalgebra is diagonal with trace zero, meaning we need only consider the case where $H_d$ is diagonal. Let $H_d = \text{diag} \{ \lambda_1, \lambda_2, \ldots, \lambda_n\}$, where the $\lambda_i$ are the eigenvalues of $H_d$. 

We choose the basis of ${\mathfrak{k}}$ to be the set of $n \times n$ matrices given by $B_{ij} = |e_i\rangle \langle e_j| - |e_j\rangle \langle e_i |$, $i<j$, where $|e_i\rangle$ is the column vector with a 1 in row $i$ and zero everywhere else. The size of this basis is $\text{dim}\, {\mathfrak{k}} = n(n-1)/2$. 

The commutator of $H_d$ with the basis elements of ${\mathfrak{k}}$ is given by
\begin{equation}
 \big[ H_d, B_{ij} \big] = (\lambda_i - \lambda_j) \left( |e_i\rangle \langle e_j| + |e_j\rangle \langle e_i |\right),
\end{equation}
demonstrating that to ensure ${[} H_d, B_{ij} {]} = 0$ we require $\lambda_i = \lambda_j$. This means that $\dim(\{A\in {\mathfrak{k}} \,|\,[H_d,A]=0)$ is given by the number of pairs $M$ of eigenvalues of $H_d$ that are degenerate, giving $\dim\mathcal O(iH_d) = \text{dim}\, {\mathfrak{k}} - M$. 

So, for example, if all eigenvalues are distinct, $M=0$ meaning $\dim\mathcal O(iH_d) = \text{dim} \, {\mathfrak{k}}$. If all eigenvalues are identical, then $M = \frac{1}{2} n (n-1) =  \text{dim} \, {\mathfrak{k}}$ meaning $\dim\mathcal O(iH_d) = 0$. 

This shows that the more eigenvalues that are degenerate, the smaller the chance the the adjoint orbit allows us to choose a direction that makes the bound tight.

As an example, consider the case $\text{SU}(2)/\text{SO}(2)$ discussed in the previous Section. Here $d_k = 1$, $d_p = 2$, so equality in Eq.~(\ref{eqDimensionCountingCrit}) is achieved when the two eigenvalues of $H_d$ are not degenerate. Specifically, in this case the adjoint orbit is one-dimensional, and since $G/K$ is the two-sphere, this single degree of freedom for the adjoint orbit suffices to choose arbitrary directions on the two-dimensional manifold, meaning achieving a minimal geodesic from the origin to the diameter is always possible.

\section{Examination of the tightness of our bounds with Cartan controls}
\label{secQSLWithCartanControls}

In the previous Section we developed a criterion that was {\emph{sufficient}} to show our speed limit bounds were tight, based on determining the dimension of the adjoint orbit. 
As this criterion is not {\emph{necessary}}, however, this Section examines what else can be said about the tightness of the bounds.
We do this mostly for the controllable case $\mathfrak g=\mathfrak{su}(n)$ by using the root system of $({\mathfrak{g}}, {\mathfrak{k}})$, and we illustrate the approach using ${\mathfrak{k}} = {\mathfrak{so}}(n)$.


We begin by considering the symmetric spaces described in the previous section as arising from the situation where the controls form a Cartan decomposition of the full Lie algebra. 
As before, the control algebra is denoted ${\mathfrak{k}}$ and the associated control group denoted $K = e^{\mathfrak{k}}$. This decomposition is often used in quantum control problems. 
The main point is that a Cartan decomposition provides a decomposition of the full Lie algebra of the form ${\mathfrak{g}} = {\mathfrak{p}} \oplus {\mathfrak{k}}$ with ${\mathfrak{p}} = {\mathfrak{k}}^\perp$, that satisfies the relations
\begin{eqnarray}
{[}{\mathfrak{k}}, {\mathfrak{k}}{]} & \subseteq & {\mathfrak{k}}, \label{eqSymmetricCondition1} \\
{[}{\mathfrak{k}}, {\mathfrak{p}}{]} & \subseteq & {\mathfrak{p}}, \label{eqSymmetricCondition2}  \\
{[}{\mathfrak{p}}, {\mathfrak{p}}{]} & \subseteq & {\mathfrak{k}}. \label{eqSymmetricCondition3}
\end{eqnarray}
These conditions include those required for a reductive space, plus the additional condition (\ref{eqSymmetricCondition3}). 
Here the Lie algebra is again equipped with the inner product \eqref{eqKillingFormSUN} in order to match the speeds and manifold diameters computed in the previous section.

There are precisely three Cartan decompositions of ${\mathfrak{su}}(n)$ \cite{dalessandro2008}. They are ${\mathfrak{k}} = {\mathfrak{so}}(n)$, ${\mathfrak{k}} = {\mathfrak{sp}}(\frac{n}{2})$, and $\mathfrak k = \mathfrak s(\mathfrak u(p)\oplus\mathfrak u(q))$ with $p+q=n$, where 
\begin{multline}
    \mathfrak s(\mathfrak u(p)\oplus\mathfrak u(q)) \\= \bigg\{\bigg(\begin{tabular}{cc} A & 0 \\ 0 & B \end{tabular}\bigg)\,\bigg|\, A\in \mathfrak u(p), B\in \mathfrak u(q), \mathrm{Tr} A =-\mathrm{Tr} B\,\bigg\}.
    \label{eqSuPQAlg}
\end{multline}
These three decompositions are associated with the three possible symmetric spaces of $\text{SU}(n)$ we met before.

To proceed we need the following notion: 
A Cartan subalgebra of $\mathfrak g$ (with respect to a Cartan decomposition $\mathfrak g = \mathfrak p \oplus\mathfrak k$) is a maximal abelian subalgebra $\mathfrak a$ contained in $\mathfrak p$ \cite{dalessandro2008} (subalgebras contained in $\mathfrak p$ are abelian because of Eq.\ \eqref{eqSymmetricCondition3}).
All Cartan subalgebras are conjugate via an element $k\in K$ and every element of $\mathfrak p$ is contained in a Cartan subalgebra \cite{dalessandro2008}. 
In particular, for every $X\in \mathfrak p$ there are $k \in K$ and $A\in \mathfrak a$ so that
\begin{equation}
    X = k A k^{-1}.
\label{eqAConjugateToP}
\end{equation}
From now on we assume that ${\mathfrak{g}} = {\mathfrak{su}}(n)$.
It is possible to use the maximal tori theorem to show \cite{khaneja2001} that the fastest way to generate any target unitary $U_{\mathrm{targ}}$ is to find the the smallest $\tau$ such that it is possible to write
\begin{equation}
U_{\text{targ}} = k_1 \exp(v \tau) k_2
\label{eqMaximalToriSolution}
\end{equation}
with $k_1, k_2 \in K$ and $v\in \mathfrak p$ of the form
\begin{equation}
v = \sum_{i=1}^m \beta_i X_i,\quad \beta \geq 0,\, \sum \beta_i =1,\, X_i \in \mathcal{W}(iH_d) 
\label{eqMaximalToriSolutionForV}
\end{equation}
where $\mathcal{W}(iH_d) = {\mathfrak{a}} \cap {\mathcal{O}(iH_d)}$ is the Weyl orbit of $iH_d$.
Note that Eq.~(\ref{eqMaximalToriSolution}) does not actually give a specific minimal time solution; it merely states the form it must take, and reduces the difficulty of the (usually numerical) optimization problem. 

Clearly $v\in{\mathfrak{p}}$ and gives the direction of the geodesic connecting the identity and $U_{\text{targ}}$ in $G/K$, so (\ref{eqMaximalToriSolution}) shows the the correct control strategy is to apply strong controls initially to pick the correct direction in $G/K$ provided the adjoint orbit allows the direction, drift for a time with all controls at zero, then apply strong controls again to move to the final desired $U_{\text{targ}}$ within the coset.

If $v$ lies in ${\mathcal{O}}(iH_d)$, we can generate it and will always be capable of moving on a geodesic between any two points in $G/K$, including from the identity to the point the furthest away corresponding to the diameter of $G/K$. 
Since all elements of the Weyl orbit commute, $\exp(v \tau)$ can be written
\begin{equation}
\exp(v \tau) = \exp(\beta_1 X_1) \exp(\beta_2 X_2) \ldots \exp(\beta_m X_m)
\label{eqPiecewiseExponentialPropagator}
\end{equation}
with $\beta_i$ and $X_i$ as in \eqref{eqMaximalToriSolutionForV}.
Because the elements of the Weyl orbit $\mathcal W(iH_d)$ are a subset of the adjoint orbit ${\mathcal{O}(iH_d)}$, we are clearly capable of implementing $\exp(v \tau)$ through the action of the drift and arbitrarily strong controls.

It is important to note, however, that the fastest way of implementing a unitary by using the available controls, i.e.\ the path described by (\ref{eqPiecewiseExponentialPropagator}), is not necessarily a minimal geodesic between the initial and final points even though it is a piecewise geodesic. 
Only if the right hand side of (\ref{eqPiecewiseExponentialPropagator}) consists of single exponential is it possible that the time this fastest path takes coincides our lower bound given by Eqs.~(\ref{eqSpeedLimitSOEven}) -- (\ref{eqSpeedLimitUpUq}).

We now examine the question as to when the $v$ in Eq.~(\ref{eqPiecewiseExponentialPropagator}) lies within the adjoint orbit, making our speed limit lower bounds tight.

As said in the previous section, the fact that $G/K$ is homogeneous implies that $\mathrm{diam}(G/K) = \sup_{x\in G/K}d(o,x)$, meaning we need only look for the point $x$ corresponding to the target unitary that is furthest from the group identity along a single-parameter geodesic.
This point has the property that a geodesic starting at the origin stops to be length minimizing after running through $x$.
The set of points where geodesics starting at the origin $o$ stop to be length minimizing is known as the cut locus (of the origin).
By the Hopf–Rinow theorem \cite{gallier}, there is for every $x\in G/K$ a minimizing geodesic joining it with the origin.
If $G$ is simply connected (as is the case for $\mathrm{SU}(n)$), the symmetric space $G/K$ is also simply connected \cite{helgason1962} (Proposition 3.6) which implies that the cut locus coinicides with what is called the first conjugate locus \cite{gorodski} (Theorem 3.5.4).

The conjugate locus can be described in terms of the positive roots $\Delta^+ (\mathfrak{g}, {\mathfrak{a}})$ of the Lie algebra ${\mathfrak{g}}$ with respect to the Cartan subalgebra ${\mathfrak{a}}$. Specifically, the exact form of the conjugate locus of a point $x\in M$ is given by \cite{yang2007, crittenden1962, gorodski}:
\begin{equation}    \label{eqConjugateLocusDefinition}
    C(x) = \big\{x \cdot e^A k \,\big| \,k\in K,\,A\in\mathfrak a\ \text{s.t.\ \eqref{eqConjugateConditionLambda} holds}\big\}
\end{equation}
with the root condition
\begin{equation}\label{eqConjugateConditionLambda}
    \exists \alpha\in\Delta^+(\mathfrak g,\mathfrak a), \, 0\ne m\in\mathbb Z : \, \alpha(A) = im\pi
\end{equation}
and the first conjugate locus corresponds to $m = \pm 1$.
Note that the locus $C(x)$ is $K$-invariant (i.e.\ invariant under the right action by elements of $K$) and invariant under inversion.
As explained previously, we only care about $C(o)$. In this case we have $o\cdot e^A k = [e^{A}k]=[\exp(k^{-1}Ak)]$ so that the conjugate (and, hence, cut) locus consists precisely of the points $[\exp\mathcal O(A)]$ where $A$ satisfies the root condition \eqref{eqConjugateConditionLambda}.

In order to illustrate this approach, we consider the simplest cases: $\text{SU}(2) / \text{SO}(2)$ and $\text{SU}(3)/\text{SO}(3)$.

For $\text{SU}(2) / \text{SO}(2)$ we consider the control problem (\ref{eqHSU2ControlProblem}). 
The control algebra is $\mathfrak k = \text{span} \{ i\sigma_x \}$, and the Cartan subalgebra is given by $\mathfrak a = \mathrm{span}\{i\sigma_z\}$. 
The locus $C(o)$ consists of points $[e^{i\eta\sigma_z}]\cdot e^{i \beta\sigma_x}$ where $\beta\in[0,2\pi]$ and $A=i\eta\sigma_z$ must satisfy the root condition. 
There is a single positive root $\alpha_1$ in this case given by $\alpha_1(i \eta \sigma_z) = 2 i \eta$. 
This means we require $ 2 i \eta = \pm i \pi$. 
The cut locus is therefore given by the set $\{[e^{\pm i\sigma_z \pi/2} e^{i\beta\sigma_x}]\,|\,\beta\in[0,2\pi]\}$ which actually only contains the single coset $[e^{i\sigma_z\pi/2}]$ because all $e^{\pm i\sigma_z \pi/2}e^{i\beta\sigma_x}$ determine the same coset.
A geometrical way to understand this is that the coset $[e^{i\sigma_z \pi/2}]$ is the unique point on $S^2=\mathrm{SU}(2)/\mathrm{SO}(3)$ that is antipodal point to the origin. 
The group action of $\mathrm{SO}(2)$ acts on the sphere by rotating about the axis going through the origin and hence fixes this antipodal point.
Since the control elements $k_1$ and $k_2$ can be applied arbitrarily quickly, our drift will hit the conjugate locus at time $t=\pi/2$ giving the expected speed limit, and showing the bound is tight.

$\text{SU}(3)/\text{SO}(3)$ is more complex. Generally speaking, for an Riemannian manifold the set of points corresponding to the diameter and the set of points corresponding to the cut locus are not the same. It is the case, however, that the diameter locus must be a (possibly equal) subset of of the cut locus.

Consider the question as to whether there is single-parameter geodesic that lies in the Weyl orbit that, up to conjugation by the controls, lies on the conjugate locus at a specific time $t_{\text{qsl}}$ given by the quantum speed limit bound in Eqs.~\eqref{eqSpeedLimitSOEven} -- \eqref{eqSpeedLimitSOOdd}. That is, whether for a given drift $H_d$ we can find a solution for a specific $A \in {\mathfrak{a}}$, $k_1, k_2 \in K$ satisfying
\begin{equation}
\exp[A] = k_1 \exp[iH_d \, t_{\text{qsl}}] k_2
\label{eqDiameterNecessaryNotSufficient}
\end{equation}
with $\alpha(A) = \pm i\pi$. If we can find such an $A$, then we know we can move to the cut locus on a single minimal geodesic, but this final point may not lie on the set of diameter points. If it does, our bound is clearly tight, since in order to reach the diameter, our geodesic must fail to be distance minimizing for the first time at that point. This means that the condition given by  Eq.~(\ref{eqDiameterNecessaryNotSufficient}) is necessary but not sufficient. To make it sufficient, it would be necessary to be able find a solution to Eq.~(\ref{eqDiameterNecessaryNotSufficient}) for {\emph{all}} $A\in {\mathfrak{a}}$, which is generally not possible.



The Cartan subalgebra ${\mathfrak{a}}$ has rank two, and can be parameterised as $A = c_1 h_1 + c_2 h_2$ where $c_1, c_2$ are real parameters and we use the Cartan-Weyl basis
\begin{equation}
h_1 = i\begin{pmatrix} 1 & 0 & 0\\ 0 & -1 & 0 \\ 0 & 0 & 0 \end{pmatrix}, \,\,\,\,\,\,
h_2 = i\begin{pmatrix} 0 & 0 & 0\\ 0 & 1 & 0 \\ 0 & 0 & -1 \end{pmatrix}.
\end{equation}
There are three positive roots for ${\mathfrak{su}}(3)$ and their action on $A$ is given by 
\begin{eqnarray}
\alpha_1(c_1 h_1 + c_2 h_2) &=& i(2c_1 - c_2) 
\label{eqRootActionsOnCartanWeylBasis1} \\
\alpha_2(c_1 h_1 + c_2 h_2) &=& i(-c_1 + 2 c_2) 
\label{eqRootActionsOnCartanWeylBasis2} \\
\alpha_3(c_1 h_1 + c_2 h_2) &=& i(c_1 + c_2).
\label{eqRootActionsOnCartanWeylBasis3}
\end{eqnarray}
Applying these roots to (\ref{eqConjugateConditionLambda}) shows that the $A\in\mathfrak a$ generating the conjugate locus can be parameterized as the union of the three sets of Lie algebra elements
\begin{eqnarray}
A_1 &=& i \, \text{diag} \{ c_1, \, c_1-m_1\pi, \,-2 c_1 +m_1\pi\} \nonumber \\
A_2 &=& i \, \text{diag} \{ c_1, \, \frac{1}{2} (-c_1 + m_2 \pi), \, -\frac{1}{2}(c_1 + m_2\pi) \} \nonumber \\
A_3 &=& i \, \text{diag} \{ c_1, \, -2c_1 + m_3\pi, \, c_1 - m_3\pi\}.
\label{eqSO3LocusGenerators}
\end{eqnarray}
where $c_1$ is an arbitrary real parameter, and $m_i=\pm$.

This shows that the bound (\ref{eqSpeedLimitSOOdd}) will be exact if we can find integers $m_i$ not equal to zero and  control group elements $k, k'$, such that
\begin{equation}
k \exp[A] k' = \exp[i H_d \, t_{\text{qsl}}]
\end{equation}
for all 
$A$ satisfying (\ref{eqSO3LocusGenerators}), as this ensures we can reach the entire cut locus, of which the diameter is a subset. Since each $k\in K$ has three parameters, this is already an eight-parameter problem, and is analytically difficult. Higher dimensional groups will pose an even bigger problem.

We can, however, gain some partial information by making use of Eqs.~(\ref{eqRootActionsOnCartanWeylBasis1}) -- (\ref{eqRootActionsOnCartanWeylBasis3}). We note that any drift $iH_d \in {\mathfrak{p}}$ can be moved into a Cartan subalgebra ${\mathfrak{a}}$ via conjugation by some controls, i.e.\ $i H_d^{\mathfrak{a}} = k iH_d k^\dagger$. This is a unitary transformation which does not change the spectrum, and since the Cartan subalgebra is spanned by the real diagonal matrices with zero trace, we write $H_d^{\mathfrak{a}}$ in terms of its eigenvectors: $H_d^{\mathfrak{a}} = \text{diag} \{ \lambda_1, \lambda_2, \lambda_3 \}$ where $\lambda_3 = -\lambda_1 - \lambda_2$ since $\text{Tr} (H_d^{\mathfrak{a}}) = 0$.


If we apply Eqs.~\eqref{eqRootActionsOnCartanWeylBasis1}, \eqref{eqRootActionsOnCartanWeylBasis3} and (\ref{eqConjugateConditionLambda}) to $A = i H_d^{\mathfrak{a}}t = i t \, \text{diag}\{ \lambda_1, \lambda_2, -\lambda_1-\lambda_2 \}$ we see that to intersect the cut locus we need one of
\begin{eqnarray}
(\lambda_1 + 2\lambda_2) t &=& m_1 \pi \label{eqEvalCondition1} \\
(\lambda_1 - \lambda_2) t &=& m_2 \pi \label{eqEvalCondition2}
\end{eqnarray}
to be satisfied. 
Since $t$ is a continuous positive parameter, these conditions will almost always be satisfied for some $t$, unless $\lambda_1$ and $\lambda_2$ are chosen to make the left hand side of one of \eqref{eqEvalCondition1}, \eqref{eqEvalCondition2} equal to zero. This occurs if $\lambda_1 = \lambda_2$, or if $\lambda_1 = -2\lambda_2$. However these two conditions are equivalent, since if $\lambda_1 = -2\lambda_2$, then $\lambda_3 = -\lambda_1 - \lambda_2 = \lambda_2$ due to the zero trace condition, showing that $\lambda_1$ and $\lambda_3$ are degenerate. Consequently if any two eigenvalues are degenerate, one of \eqref{eqEvalCondition1}, \eqref{eqEvalCondition2} cannot be met.

Note that this condition does not guarantee there is no element of the Weyl orbit that produces a single-parameter geodesic from the identity to the point corresponding to the diameter, but it reduces the possibility since it ensures there is a portion of the cut locus that cannot be reached. This is because each root condition corresponds to a geodesic that intersects a different portion of the cut locus, so failing one of the conditions \eqref{eqEvalCondition1}, \eqref{eqEvalCondition2} will only result in the bound not being tight if the diameter lies on that portion of the cut locus. 

However this is more powerful than might first be imagined, since if {\emph{any}} drift Hamiltonian with degenerate eigenvalues exceeds the lower bound on the speed limit, then {\emph{all}} drift Hamiltonians with degenerate eigenvalues will exceed the lower bound. This is because the ordering of the elements of a diagonal matrix can be arbitrarily switched by controls, and multiplying the drift by a scalar does not change whether bound is tight; it merely stretches the timescale. This means all drifts with two degenerate eigenvalues have the same behaviour regarding whether the bound is tight. If this can be determined for a single case in $\text{SU}(3)/\text{SO}(3)$, the behaviour of all drift Hamiltonians is known. This is one of the questions that will be investigated numerically in the next Section.



\section{Numerical tests of the analytic speed limits}
\label{secNumericalTests}

In Section \ref{secQSLDiamAndDrift} we derived lower bounds on the quantum speed limit for various types of controls, and in Section \ref{secQSLWithCartanControls} we looked at evidence for when these bounds might be saturated, i.e.\ when the bound is actually exact.

We now examine these systems to determine the speed limit via a numerical optimization procedure. The motivation is to provide checks on both the analytic bounds as well as to test our dimension counting and eigenvalue degeneracy arguments for bound tightness laid out in Sections \ref{secBoundTightnessDimensionCounting} and \ref{secQSLWithCartanControls}. In addition, bilinear optimal control problems are seldom analytically tractable and are usually approached numerically, so our analytic results provide a ideal test for checking the performance of various optimization strategies.

Our approach was to determine the quantum speed limit numerically for a variety of drifts and a variety of Hilbert space dimensions, assuming the controls Hamiltonians generate one of the three Lie algebras ${\mathfrak{so}}(n)$, ${\mathfrak{sp}}(\frac{n}{2})$, or $\mathfrak s(\mathfrak u(p)\oplus\mathfrak u(q))(n)$. To do this we chose a series of Haar-random unitary targets, and attempted to numerically find optimal controls that, for a specific drift Hamiltonian $H_d$, would achieve that unitary at a specific chosen time $T$. That time was divided into $N$ discrete intervals ("time slots"), with the width of each time slot given by $T/N$, and the controls were assumed to have a constant amplitude over each interval, i.e.\ the controls were time-dependent but piecewise constant. In the limit of a large number of time slots, arbitrary control functions are well approximated. Specifically, we solved
\begin{equation}
i \frac{d}{dt} {U}(t) = \bigg( H_d + \sum_{j=1}^m f_j(t) H_j \bigg) U(t)
\label{eqSE2}
\end{equation}
with an initial random guess at the amplitudes in each time slot for each independent control function $f_j(t)$. We used QuTiP's optimal control package \cite{qutip} with a gradient ascent algorithm to find the control functions that maximized the overlap between the final unitary resulting from the evolution of (\ref{eqSE2}) and the desired target unitary, as given by the phase-insensitive fidelity measure
\begin{equation}
F = \frac{1}{n}\big|\text{Tr}\big[ U_{\text{target}}^{\dagger} U(T) \big] \big|.
\end{equation}

\begin{figure}[htp]
\includegraphics[width=1.0\columnwidth]{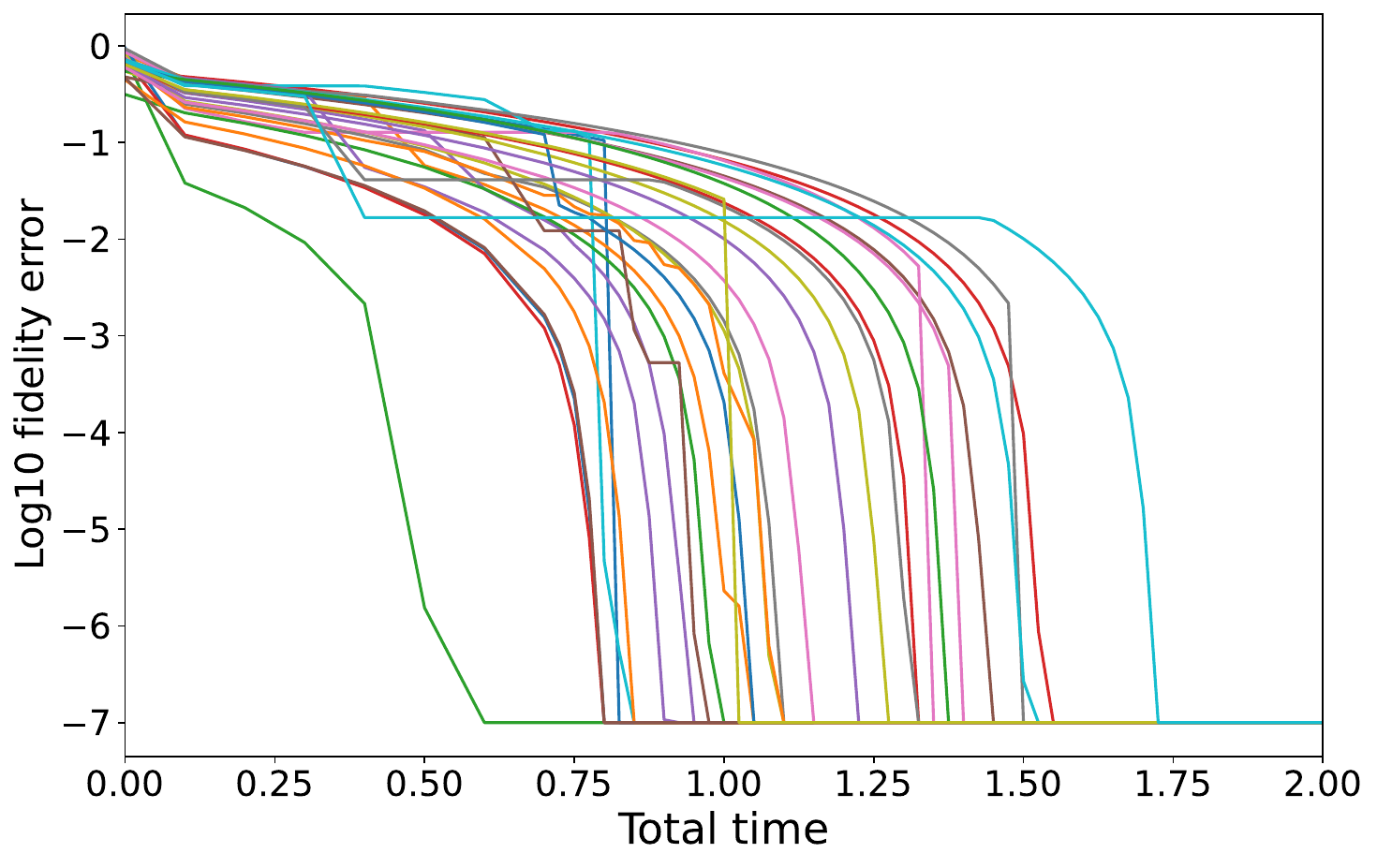}
\caption{Example of how the quantum speed limit is determined numerically, for the case with $\text{SO(3)}$ controls. Each line corresponds to a random target unitary in $\text{SU(3)}$. We attempt to find a solution for time dependent controls for a given fixed time total time $T$ (horizontal axis), and a specified drift $H_d$. As $T$ is increased, better solutions can be found, giving a better fidelity overlap with the target unitary. When the fidelity error is lower than some cutoff, we take this to mean we have found a solution for the control pulse that can generate the unitary. If this is repeated many times for many random unitary targets, the speed limit is taken to be the time for which we can find a control pulse for all possible targets in this time or less. This plot shows 30 Haar random unitary targets with $H_d = \text{diag} \{ 1,0,-1\}$ and 100 time slots. }
\label{fig:waterfall_SO3_50targets}
\end{figure}

This process was then repeated many times with different random initial guesses to help the optimizer becoming stuck in local minima. For each target unitary, we gradually increased the time $T$ until a solution could be found where the fidelity error $1-F$ was less than a cutoff of $10^{-7}$. This was repeated for a large number of random unitaries, and the quantum speed limit for that particular drift was taken to be the lowest time for which we could guarantee solutions for all the unitaries with a fidelity error less than the cutoff. This is illustrated in Figure~\ref{fig:waterfall_SO3_50targets} with a small sample of the results for the $\text{SU(3)}/\text{SO}(3)$ case for a particular drift corresponding to a predicted analytic quantum speed limit of $t_{\text{qsl}} = 1.81$. It shows how the fidelity error for any given target reduces as more time is allowed, until we reach a sudden drop in the error which we interpret as the existence of a set of control functions that can achieve that unitary. 

Not all drift Hamiltonians $H_d$ need to be examined. First, if $iH_d$ has some overlap with ${\mathfrak{k}}$ then this portion can be removed arbitrarily quickly by application of the controls, so we can assume $i H_d \in {\mathfrak{p}}$. Second, since $i H_d \in {\mathfrak{p}}$, due to (\ref{eqAConjugateToP}) it can also be moved into a subspace of ${\mathfrak{p}}$ corresponding to a Cartan subalgebra ${\mathfrak{a}}$ by application of the controls. This means we need only consider drift Hamiltonians drawn from ${\mathfrak{a}}$ (multiplied by $i$).

There are a number of reasons that the numerical approach may provide a speed limit higher than the true one, making it difficult to determine if the lower bounds given by Eqs.~(\ref{eqSpeedLimitSOEven}) -- (\ref{eqSpeedLimitUpUq}) are truly tight. First, for a given time there may have been a better solution that the optimizer simply missed, even with many attempts with random initial conditions. Second, because we have divided the total time $T$ into $N$ time slots, elements of the control group cannot be performed arbitrarily fast; they take at least $T/N$. Both of these serve to ensure the speed limit found numerically will be slightly higher than the true speed limit. Third, the since the testing is done with a set of discrete choices of time $T$, there may be a fast solution at a specific low $T$ that we don't see because that value of $T$ is not tested, giving the illusion that the speed limit for that unitary is higher than it actually is. Conversely, we draw the target unitaries from a Haar-random set. As the dimension of the Hilbert space increases, it becomes increasingly difficult to properly sample the set of possible unitaries, and this is exacerbated by the fact that higher dimensions take longer to simulate so fewer targets can be sampled.

With these caveats in mind, we now examine the results of the numerical optimization process. We first consider the case where the controls generate the $\text{Sp}(n/2)$ subgroup. As discussed in Section~\ref{secQSLWithCartanControls}, from dimension counting arguments we might expect the speed limit bounds given by (\ref{eqSpeedLimitSpEven}) and (\ref{eqSpeedLimitSpOdd}) to be tight. The elements of the Lie algebra ${\mathfrak{sp}}(\frac{n}{2})$ have the form 
$$\begin{pmatrix}L_1&L_2\\ -L_2^*&L_1^*\end{pmatrix}$$
with $L_1$ skew-Hermitian and $L_2 =L_2^T$, where $L_1$, $L_2$ are complex and $\frac{n}{2} \times \frac{n}{2}$ in size. One chooses a basis for this space, and the control Hamiltonians will be given by this basis multiplied by $i$.  

\begin{figure}[htp]
\includegraphics[width=1.0\columnwidth]{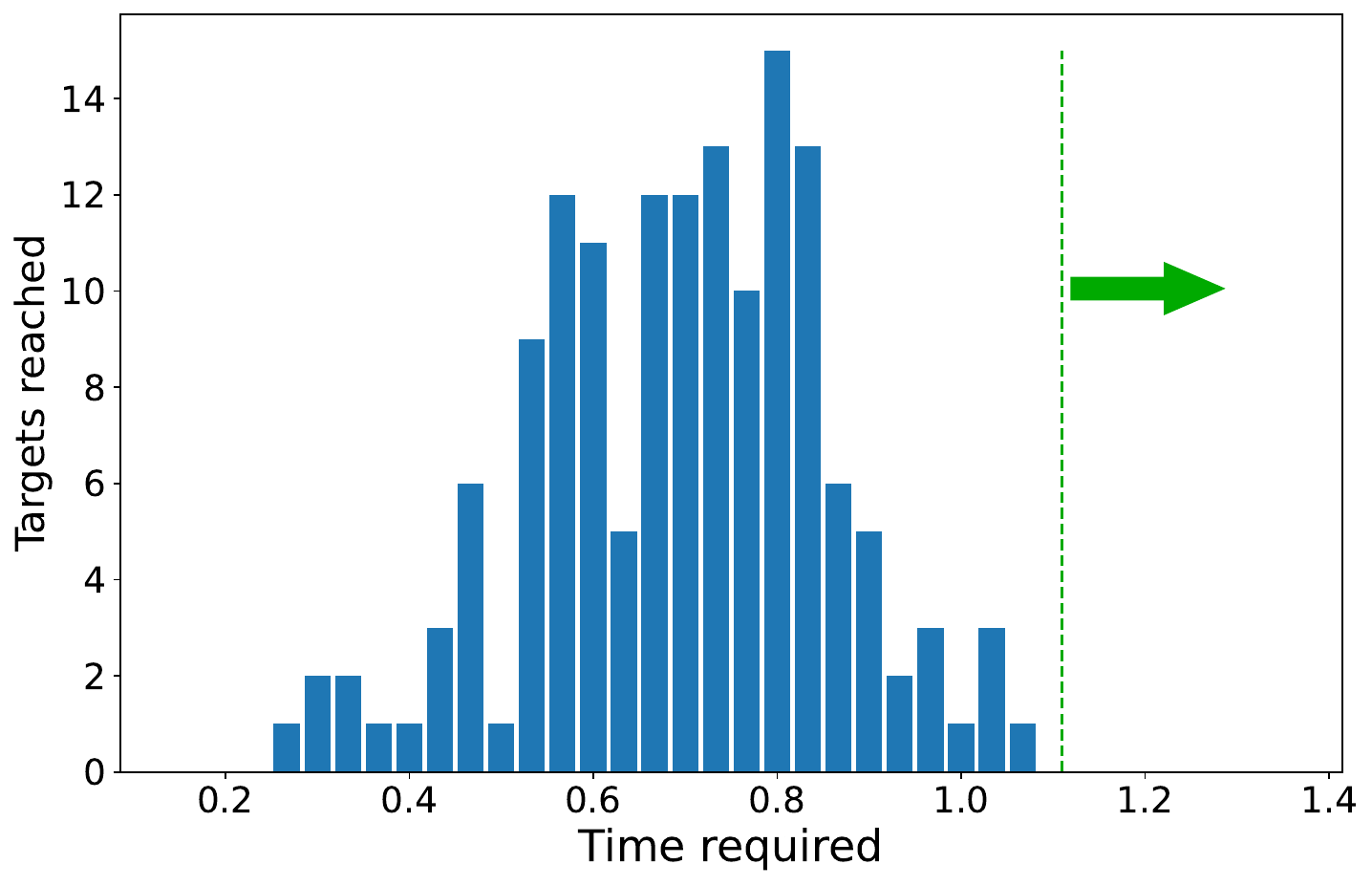}
\caption{Speed limit for the $\text{SU}(4)/\text{Sp}(2)$ case, with a drift $H_d = \text{diag}\{1, -1, 1, -1 \} $. The histogram shows the fastest possible times to achieve 150 randomly chosen unitary targets when using $\text{Sp}(\frac{n}{2})$ controls, with the analytic lower bound given by (\ref{eqSpeedLimitSpEven}) represented by the vertical green line. As all targets can be met in a time less than the bound, and some targets are at the bound, the bound is tight.}
\label{fig:speedlimit_SU4xSp2_1_-1_1_-1}
\end{figure}

As discussed above, we need only consider drift Hamiltonians that lie within the Cartan subalgebra, which drastically reduces the possibilities. For ${\mathfrak{sp}}(\frac{n}{2})$ this is given by  matrices of the form \cite{dalessandro2008}
\begin{equation}
A = \begin{pmatrix} D & 0 \\ 0 & D \end{pmatrix}
\end{equation}
with $D$ diagonal and $D \in {\mathfrak{su}}(\frac{n}{2})$. Figure~\ref{fig:speedlimit_SU4xSp2_1_-1_1_-1} shows results for the $\text{SU}(4)/\text{Sp(2)}$ case, with a drift Hamiltonian $H_d = \text{diag} \{ 1,-1,1,-1 \}$. Up to a constant factor, this is in fact the only drift Hamiltonian that lies within the Cartan subalgebra. As expected, all random target unitaries chosen can be reached with a time under the speed limit given by (\ref{eqSpeedLimitSpEven}), and the maximal time falls on the speed limit, showing that the bound is tight.


\begin{figure}[htp]
\includegraphics[width=1.0\columnwidth]{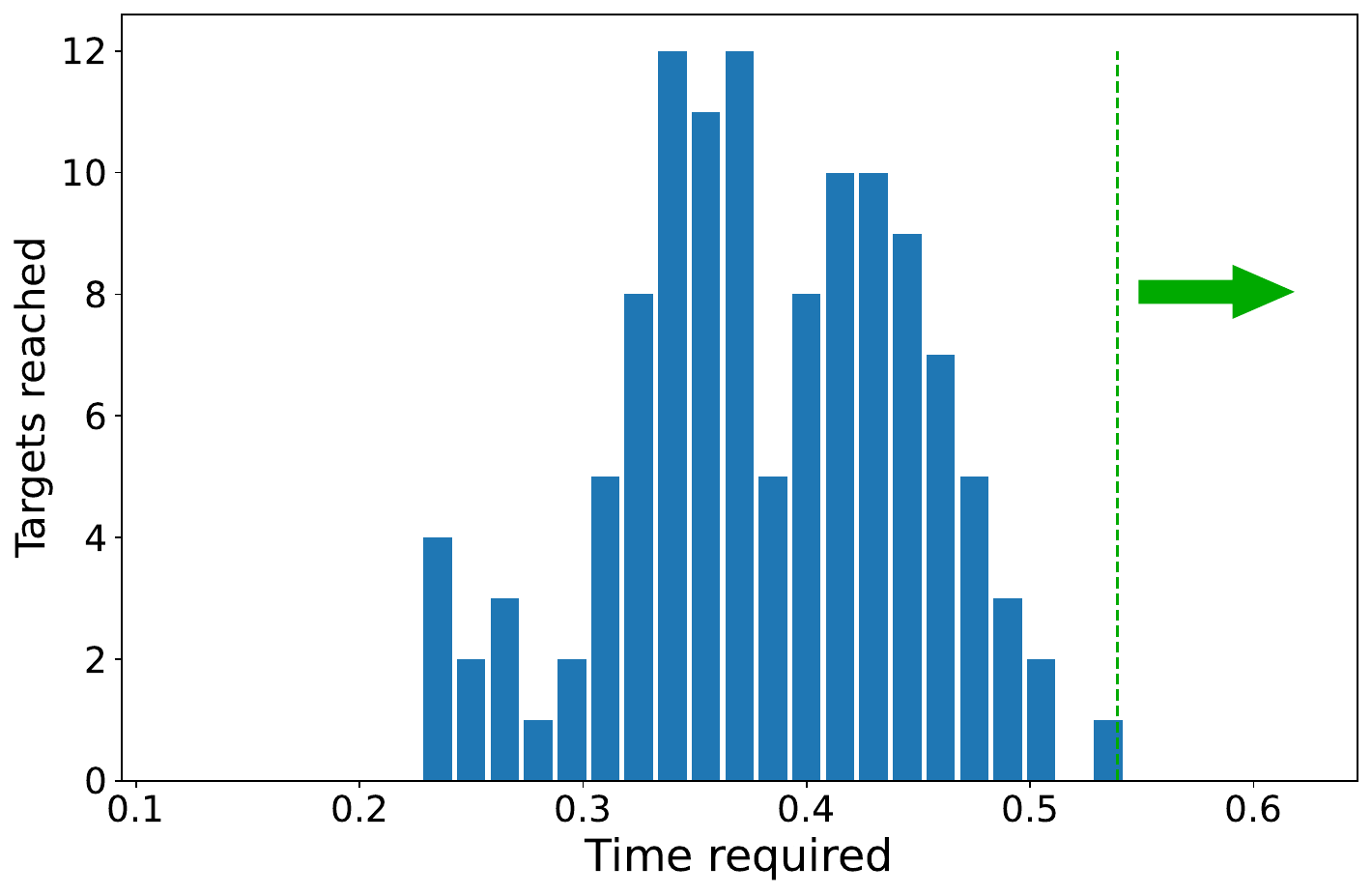}
\caption{Speed limit for the $\text{SU}(5)/\text{S}(\text{U}(2) \times \text{U}(3))$ case, with a drift Hamiltonian given by (\ref{eq14Hamiltonian}). The histogram shows the fastest possible times to achieve 120 randomly chosen unitary targets when using $\text{S}(\text{U}(2) \times \text{U}(3))$ controls, with the analytic lower bound given by (\ref{eqSpeedLimitUpUq}) represented by the vertical green line. As all targets can be met in a time less than the bound, and some targets are at the bound, the bound is tight.}
\label{fig:speedlimit_SU5xU2U3_14}
\end{figure}

Next we consider the case where the controls generate the $\text{S} (\text{U}(p) \times \text{U}(q))$ subgroup of $\text{SU}(n)$, with $p+q=n, p\leq q$. 
Its Lie algebra $\mathfrak s(\mathfrak u(p)\oplus\mathfrak u(q))$ is given by Eq.~\eqref{eqSuPQAlg}.
Again, from dimension counting arguments one would expect the bound given by (\ref{eqSpeedLimitUpUq}) to be tight. 
The Cartan subalgebra is given by matrices of the form \cite{dalessandro2008}
\begin{equation}
A = \begin{pmatrix} 0 & B \\ -B^T & 0 \end{pmatrix}
\end{equation}
where $B$ is a real $p \times q$ matrix that is zero everywhere except for the first $p$ columns, which is given by a $p\times p$ diagonal matrix. We chose our drift Hamiltonian to be given by
\begin{equation}
H_d = i\begin{pmatrix} 0 & 0 & 1 & 0 & 0 \\ 0 & 0 & 0 & 4 & 0  \\ -1 & 0 & 0 & 0 & 0 \\ 0 & -4 & 0 & 0 & 0 \\ 0 & 0 & 0 & 0 & 0 \\ \end{pmatrix}.
\label{eq14Hamiltonian}
\end{equation}
Figure~\ref{fig:speedlimit_SU5xU2U3_14} shows results for the $\text{SU}(5)/\text{S}(\text{U}(3) \times \text{U}(2))$ case with the drift Hamiltonian given by (\ref{eq14Hamiltonian}). As expected, all random target unitaries chosen can be reached with a time equal to or less than the speed limit given by (\ref{eqSpeedLimitUpUq}). Again, we conclude that in this case the bound is tight.

We come now to the third and final case, $\text{SU}(n)/\text{SO}(n)$. Dimension counting arguments suggest that we cannot always rely on the bound being tight, and at least in the $\text{SU}(3)/\text{SO}(3)$ case we expect the bound to fail to be tight if the drift Hamiltonian has a degenerate eigenvalue.

\begin{figure}[htp]
\includegraphics[width=1.0\columnwidth]{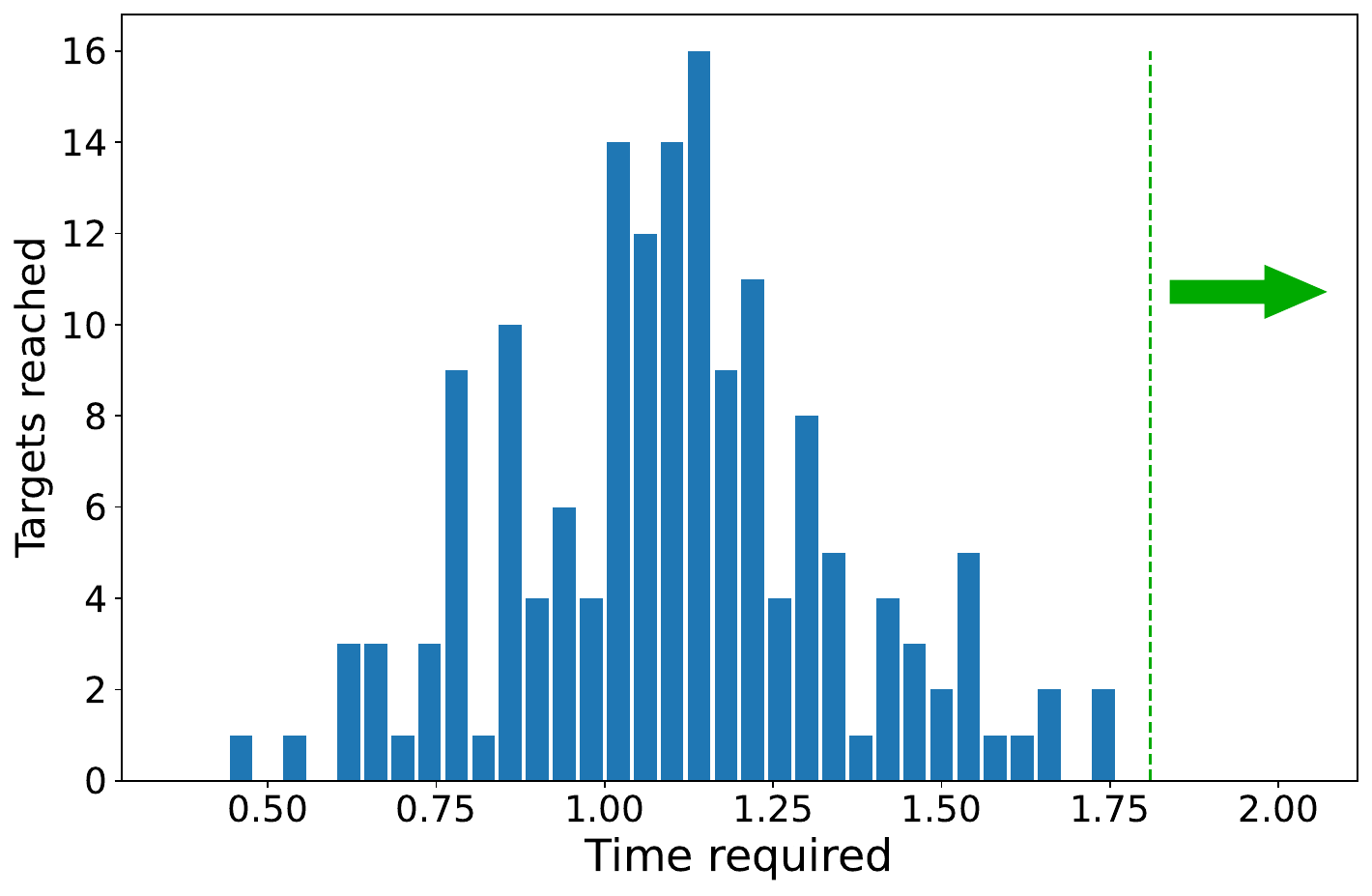}
\caption{Speed limit for the $\text{SU}(3)/\text{SO}(3)$ case, with a drift $H_d = \text{diag} \{1,0,-1 \} $. The histogram shows the fastest possible times to achieve 160 randomly chosen unitary targets when using $\text{SO}(3)$ controls, with the analytic lower bound given by (\ref{eqSpeedLimitSOOdd}) represented by the vertical green line. As all targets can be met in a time less than the bound, and some targets are at the bound, the bound is tight.}
\label{fig:speedlimit_SU3xSO3_1_0_-1}
\end{figure}

The Lie algebra ${\mathfrak{so}}(n)$ associated with the $\text{SO}(n)$ control group is the set of $n \times n$ traceless skew-Hermitian complex matrices, and the Cartan subalgebra is the set of real, diagonal, and traceless matrices. Our numerics were carried out for the $\text{SU}(3)/\text{SO}(3)$ case, where  the control Hamiltonians were given by the three Gell-Mann matrices $\lambda_2, \lambda_5$ and $\lambda_7$, and the Cartan subalgebra is spanned by $i\lambda_3$ and $i\lambda_8$. We first consider a drift Hamiltonian $H_d = \text{diag} \{ 1,0,-1\}$ which clearly does not have degenerate eigenvalues. The results are shown in Figure~\ref{fig:speedlimit_SU3xSO3_1_0_-1}. Interestingly, we see that the speed limit lower bound is still tight. No target unitary takes longer than this lower bound.

\begin{figure}[htp]
\includegraphics[width=1.0\columnwidth]{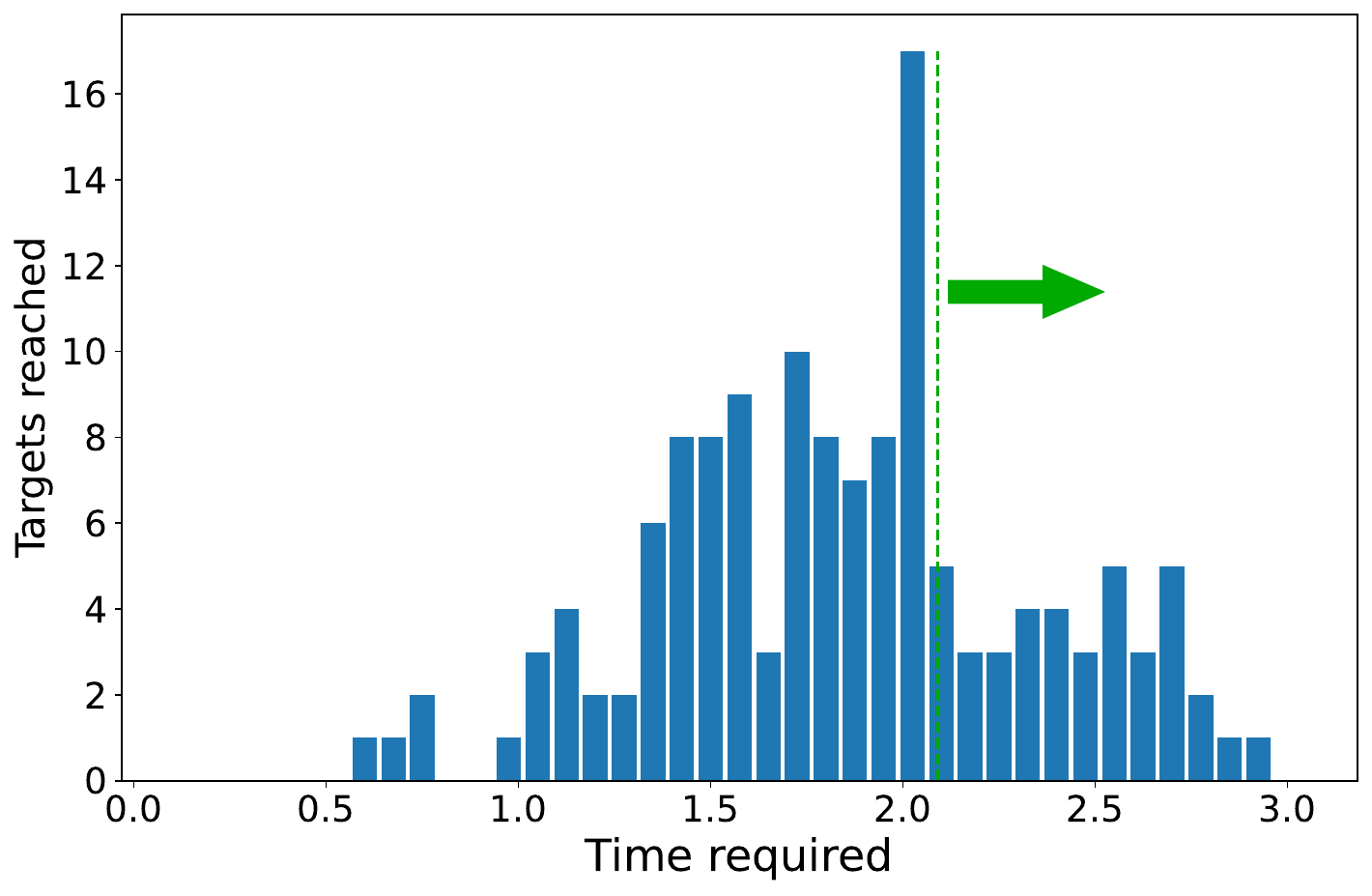}
\caption{Speed limit for the $\text{SU}(3)/\text{SO}(3)$ case, with a drift $H_d = \text{diag} \{1, -0.5, -0.5 \}$. The histogram shows the fastest possible times to achieve 140 randomly chosen unitary targets when using $\text{SO}(3)$ controls, with the analytic lower bound given by (\ref{eqSpeedLimitSOOdd}) represented by the vertical green line. Consequently the bound is not tight for this particular drift, as expected due to the fact that $H_d$ has two degenerate eigenvalues.}
\label{fig:speedlimit_SU3xSO3_1_-0.5_-0.5}
\end{figure}

Finally, we consider the case with a drift $H_d = \text{diag} \{ 1,-\frac{1}{2},-\frac{1}{2} \}$, which {\emph{does}} have a degenerate eigenvalue. The results are shown in Figure~\ref{fig:speedlimit_SU3xSO3_1_-0.5_-0.5}, and we see that while the analytic lower bound given by (\ref{eqSpeedLimitSOOdd}) is still respected it is no longer tight, which is what we expect due to $H_d$ possessing degenerate eigenvalues.

Collectively these results provide a check on the analytic results for the lower bounds on the quantum speed limit. They confirm that the bounds (\ref{eqSpeedLimitSOEven}) -- (\ref{eqSpeedLimitUpUq}) are accurate, showing that if we consider all possible unitaries, there will be at least one that takes at least this long to generate. These simulations also support our conjecture that for the $\text{Sp}(\frac{n}{2})$ and $\text{S} (\text{U}(p) \times \text{U}(q))$ control schemes, the bounds are tight, meaning that there is at least one unitary that takes exactly that long to produce, but no unitaries will take longer. Furthermore, the results show that, for the $\text{SO}(3)$ control case where the drift has a pair of degenerate eigenvalues, the bound is respected but is not tight, as expected. Interestingly, the bound with $\text{SO}(3)$ control case where the drift has distinct eigenvalues does appear to be tight, at least for the particular drift Hamiltonian we chose. 

Finally, we see that numerical optimization techniques to find optimal control pulses for quantum systems appear to work remarkably well. Optimal pulses are found that respect the analytic bounds exactly, providing evidence that such methods can be trusted for bilinear control problems.

\section{Speed limits without a full set of Lie algebra controls}
\label{secQSLWithoutFullCartan}

The previous sections have obtained lower bounds on the quantum speed limit for systems with arbitrary drifts and with controls that form a closed subgroup of $\text{SU}(n)$, as well as considering in more detail the case where the control Hamiltonians are one of the Lie algebras ${\mathfrak{so}}(n)$, ${\mathfrak{sp}}(\frac{n}{2})$, or ${\mathfrak{s}}({\mathfrak{u}}(p)\oplus {\mathfrak{u}}(q))$. The number of control Hamiltonians required to span these Lie algebras is given by $d_k$ in Table~\ref{tabDimensions0}, and can be seen to scale quadratically in $n$. Such a situation might seem to be difficult to arrange in practice.

However it is important to realise that the controls themselves need not provide a full basis for the algebra, but rather that the dynamical Lie algebra generated through repeated application of the commutators of the controls provide such a basis. Clearly, if we have a full set of controls that already provide a basis, that is enough. But the question is, how few control Hamiltonians do we actually need to generate these algebras?

It is known that the simple compact classical Lie algebras ${\mathfrak{su}(n)}$, ${\mathfrak{so}(n)}$, and ${\mathfrak{sp}(\frac{n}{2})}$ can be generated by ``one and a half'' elements \cite{bois2009}. This means that if we choose any element in the algebra, there exists a second element in the algebra that along with the first will generate the entire algebra, provided neither of the two is the identity. Consequently one never needs more than two control Hamiltonians to generate the full ${\mathfrak{so}(n)}$ or ${\mathfrak{sp}(\frac{n}{2})}$ algebras, ensuring the results in previous Sections are applicable.


Finally, so far we have only discussed systems where we have multiple control Hamiltonians, but the situation with a single control, i.e. where the system Hamiltonian is given by
\begin{equation}
H = H_d + f(t) H_c
\label{eqHwithSingleControl}
\end{equation}
is very common. It is therefore useful to derive a bound on the quantum speed limit in this case. 

Again, the full Lie algebra of the system is ${\mathfrak{g}} = {\mathfrak{su}}(n)$, and the control subalgebra is one-dimensional and is given by ${\mathfrak{k}} = \text{span} \{ i H_c \}$. This pair does not admit a Cartan decomposition unless $H_c \in {\mathfrak{so}}(2)$. Indeed, the Lie group generated by $K = \exp ({\mathfrak{k}})$ is in some cases not even topologically closed. Consequently the quotient space $G/K$ may not be a homogeneous space, let alone a symmetric space. We can, however, apply the results we derived in previous Sections to obtain a lower bound on the quantum speed limit in this case by ``embedding'' this control problem into another which does satisfy our criteria.

To obtain a bound we note that since $H_c$ is Hermitian it can be transformed into a diagonal, purely real matrix $H_c' = U H_c U^\dagger$ via a unitary transformation. In this new basis the drift is given by $H_d' = U H_d U^\dagger$. 
Changing the basis of the problem via unitary transformation cannot change the speed limit since a basis change is only a mathematical convenience. We also introduce an auxiliary control problem with the same drift $H'_d$ but with the control group given by $\text{S}(\text{U}(p)\times U(q))$ with $p+q=n$ and an associated control algebra ${\mathfrak{k}}$. This auxiliary problem {\emph{does}} admit a Cartan decomposition.

Since $i H_c'$ is diagonal, purely imaginary and traceless, it can be written
\begin{equation}
i H_c' = \begin{pmatrix} D_1 & 0 \\ 0 & D_2 \end{pmatrix}, \,\,\,\,  \text{Tr} (D_1) + \text{Tr} (D_2) = 0.
\label{eqSUNCartanDecompositionIntoUpUq}
\end{equation}
where $D_1$ and $D_2$ are diagonal, imaginary and $p \times p$ and $q\times q$ respectively. Consequently we have $D_1\in \mathfrak{u}(p), D_2\in \mathfrak{u}(q)$ and thus $i H_c' \subset {\mathfrak{k}}$. This means that the control problem
 \begin{equation}
H = H_d' + f(t) H_c'
\label{eqHWithHcPrimeControl}
\end{equation}
is the same as the auxiliary control problem, except with fewer controls. That is, it has a single control from ${\mathfrak{k}}$, rather the entire basis set of $p^2 + q^2 -1$ controls. Hence whatever the lower bound on the quantum speed limit for the auxiliary control problem, the lower bound for the system described by (\ref{eqHWithHcPrimeControl}) must be at least as large since it has a strict subset of the controls relative to the auxiliary problem. 
Since the system described by (\ref{eqHWithHcPrimeControl}) is physically equivalent to (\ref{eqHwithSingleControl}), and since the trace is unchanged by a unitary transformation, we obtain a lower bound on the quantum speed limit of (\ref{eqHwithSingleControl}) given by
\begin{equation}
T_{\text{QSL}}  \geq \frac{ \sqrt{p} \, \pi}{\sqrt{ 2 \, \text{Tr} (H_d^2)}}  \label{eqSpeedLimitUpUqSingleControl} 
\end{equation}
where we have assumed without loss of generality that $p\leq q$.

Since our split of the sizes of $D_1$ and $D_2$ in (\ref{eqSUNCartanDecompositionIntoUpUq}) is only constrained by $p\leq q$, we are free to choose the size of $p$ and $q$ to make the lower bound (\ref{eqSpeedLimitUpUqSingleControl}) as large as possible. This clearly occurs when $p=\lfloor n/2 \rfloor$, yielding 
\begin{equation}
T_{\text{QSL}}  \geq \frac{ \sqrt{\lfloor n/2 \rfloor} \, \pi}{\sqrt{ 2 \, \text{Tr} (H_d^2)}}. \label{eqSpeedLimitUpUqSingleControl2} 
\end{equation}
for the case where we have a single control. In the general case one would not expect (\ref{eqSpeedLimitUpUqSingleControl2}) to be tight, but does provide a rigorous lower bound and demonstrates how the quantum speed limit scales with dimension and how it depends on the form of the drift.

\section{Conclusion}

The purpose of this paper has been to develop a lower bound for the quantum speed limit of a controllable, finite-dimensional system, given the assumption that the controls can be arbitrarily strong. We have also investigated the circumstances under which this lower bound is not merely a bound, but is actually exact.

We have used the techniques of Lie algebras, Lie groups and differential geometry. Mindful that these areas may not be entirely familiar to many physicists, we have provided a pedagogical development of this material, making it clear why it is relevant, and constantly tying it back to the physics. We have also provided a number of examples to aid this process. Our approach has been completely general, and the basic result given by Theorem 1 holds for Hilbert spaces of arbitrary dimension, arbitrary drift Hamiltonians, and does not require specific symmetries. The only requirement is that the control group is topologically closed.

This basic result, however, does require some knowledge of the diameter of the homogeneous space corresponding to the quotient group of $\text{SU}(n)$ with the control group. While this is generally difficult to determine, exact diameters are available for symmetric spaces, allowing us to give explicit bounds in this case. It is important to note, however, that even if the exact diameter of the quotient group is not known analytically, any ability to bound the diameter, analytically or numerically, can immediately be used in our expression for the quantum speed limit, and merely results in a looser bound.

We have also examined the question of when our formula for the quantum speed limit is not merely a lower bound, but is actually exact. In the fully general case we developed a sufficiency criterion based on the dimension of the adjoint orbit and commutation relations between the drift Hamiltonian and the matrix representation of the Lie algebra corresponding to the controls. As an illustration we showed how this can be done for the case where the control group is $\text{SO}(n)$.

As this criterion for bound tightness is sufficient but not necessary, we also examined what could further be said in the case where the controls are not arbitrary, but form a Cartan decomposition of the quantum control problem. In this case bound tightness depends on the cut locus of the quotient space, which can be described in terms of the positive roots of the Lie algebras. We were not able to provide a complete statement as to when the bounds were tight, but did show how conditions on the roots would decrease the probability that the bound was tight.

Since the development of our results is somewhat abstract and mathematical, we have also examined our speed limit bounds using a numerical optimization procedure for a number of specific Hamiltonians. This purpose of this is twofold. First, it provides numerical confirmation of our explicit analytic bounds, as well as supporting our results on the link between the degree of degeneracy of the drift Hamiltonian and the tightness of the bounds. Second, it provides a general way to use numerical optimization to determine speed limits, and demonstrates that gradient descent-based techniques work well. 

Finally, we have considered the quantum speed limit in the very common quantum control case where one has a drift Hamiltonian and a single control Hamiltonian. Such a system need not meet the assumptions for our main speed limit theorem; for example the control group may not be closed, or indeed form a quotient group that is a homogeneous space. Nonetheless, we showed it is possible to embed such a problem into a group that does meet our criteria, allowing us to use our previous results and thereby provide an explicit lower bound for this case.

\vspace{1cm}

\end{document}